\RequirePackage{ifpdf}
\ifpdf\else
  \PassOptionsToPackage{dvipdfmx}{graphicx}
  \PassOptionsToPackage{dvipdfmx}{graphics}
  \PassOptionsToPackage{dvipdfmx}{xcolor}
  \PassOptionsToPackage{dvipdfmx}{hyperref}
\fi

\documentclass[twocolumn]{aastex701}

\usepackage{amsmath}
\usepackage{xurl}   % allow line breaks anywhere in long URLs (reference list)

\defcitealias{Astropy2013}{The Astropy Collaboration 2013}

\shorttitle{F-corona observed by Artemis II}
\shortauthors{Tsumura \& Arimatsu}
\begin{document}

\title{Revisiting the Optical F-corona from the Artemis II Total Solar Eclipse with RAW Data: \\ How Far Can Consumer JPEG Images Be Trusted for Astronomical Photometry?}

\author{Kohji Tsumura}
\affiliation{Frontier Research Institute for Interdisciplinary Sciences, Tohoku University, \\
6-3, Aramaki aza Aoba, Aoba, Sendai, Miyagi 980-8578, Japan}
\affiliation{Department of Natural Sciences, Faculty of Science and Engineering, Tokyo City University, \\
1-28-1, Tamazutsumi, Setagaya, Tokyo 158-8557, Japan}
\email[show]{tsumura@astr.tohoku.ac.jp}  

\author{Ko Arimatsu} 
\affiliation{National Astronomical Observatory of Japan, 2-21-1 Osawa, Mitaka, Tokyo 181-8588, Japan}
\affiliation{Ishigakijima Astronomical Observatory, 1024-1 Arakawa, Ishigaki, Okinawa 907-0024, Japan}
\email{ko.arimatsu@nao.ac.jp}

\begin{abstract}
We reanalyze the total solar eclipse images taken with a consumer camera during the Artemis II lunar flyby, using the RAW data as ground truth for our previous analysis of the released JPEG image. Calibrated with field stars, the RAW data show that the lens vignetting and the variation of the pixel solid angle reduce the signal of diffuse light by up to $\sim$70\% at the edge of the field, and that the signal on the dark side of the Moon is most likely foreground light rather than an instrumental offset. Between $10^\circ$ and $15^\circ$ elongation, the radial profiles of the F-corona follow power laws with indices of $-2.0$ to $-2.3$ (east), $-2.5$ (west), and $-2.4$ to $-2.7$ (north and south), consistent with space-based observations, and the surface brightness along the ecliptic is $\sim$26\,MJy\,sr$^{-1}$ ($\pm 7\%$) at $10^\circ$. The isophotes have flattening indices of $\sim$0.6--0.7 and superellipse indices of $\sim$1.5--1.6. Comparison with the ZodiSURF zodiacal light model supports a radial dust density index of $\alpha \simeq 1.3$ down to $\sim$0.2\,au and suggests that the vertical distribution of the dust broadens toward the Sun. The JPEG image retains nearly the full vignetting, despite its metadata, and its tone response is not a single power law; together with the zero-point treatment, these effects account for the steep and asymmetric profiles reported previously, while the flattened morphology and $\alpha \simeq 1.3$ are confirmed. We identify quantities robust in JPEG images and show that the tone response can be reconstructed from field stars.
\end{abstract}


\section{Introduction} \label{sec:intro}

\subsection{The F-corona and the Moon as a natural occulter} \label{sec:intro_fcorona}

The Fraunhofer corona (F-corona) is the inner extension of the zodiacal light,
produced by sunlight scattered by interplanetary dust particles close to the Sun
\citep{vandeHulst1947, Leinert1998}. Its brightness distribution traces the structure of the
inner zodiacal cloud. Because the F-corona must be observed against the photospheric glare,
observations have relied on coronagraphs and heliospheric imagers, such as LASCO on board SOHO
\citep{Lamy2022}, the Heliospheric Imager on board STEREO \citep{Stenborg2018}, and WISPR on
board the Parker Solar Probe \citep{Stenborg2021}, or on eclipse-like geometries in which a
planetary body blocks the solar disk \citep{Hahn2002}. The use of the lunar disk as a natural,
low-stray-light occulter dates back to the Surveyor missions \citep{Norton1967, Bohlin1971} and
has been revisited as a concept for routine coronal observations from lunar orbit
\citep{Habbal2013, Cooper2023}.

The Artemis~II crewed lunar flyby provided such a geometry. On 2026 April~7, about two hours
after closest approach, the crew photographed the totally eclipsed Sun from the Orion spacecraft
with a handheld consumer digital camera, and NASA released the image \texttt{art002e009301} as an
8-bit RGB JPEG image on the same day. Using this image, \citet[hereafter TA26]{Tsumura2026}
investigated the large-scale morphology of the optical F-corona over solar elongations of
$\sim$7$^\circ$--38$^\circ$. They reported a flattened distribution aligned with the ecliptic
(flattening indices $f = 0.52$--$0.56$; Equation~\ref{eq:flattening} in Section~\ref{sec:flattening}), radial profiles that are systematically steeper than
those derived from STEREO/HI-1A \citep{Stenborg2018}, and a radial dust density index of
$\alpha \sim 1.3$ from a comparison with the \texttt{ZodiSURF} model \citep{OBrien2026}.

\subsection{Consumer-camera JPEG images as astronomical data} \label{sec:intro_jpeg}

Images taken with consumer digital cameras are increasingly used for scientific purposes,
including measurements of night-sky brightness and artificial light at night
\citep{Hanel2018, SanchezdeMiguel2019}, photometry of bright stars \citep{Cardiel2021}, and
morphological studies in which absolute calibration is not required \citep{Lintott2008,
Krista2015}. Rendered products such as JPEG images are often the only form in which an
observation is made public, especially for opportunistic observations from crewed spacecraft.
A JPEG image, however, is a rendering of the incident flux rather than a measurement of it: the
processing chain typically includes demosaicing, white balancing, a color-space transformation
that mixes the camera channels, a tone curve, optional lens corrections, quantization to 8 bits,
and lossy compression, most of which are camera- and software-specific and poorly documented
\citep[e.g.,][]{Burggraaff2019}. Any analysis of a JPEG image must therefore assume a model for
the inverse of this chain, which usually cannot be verified from the JPEG image itself.

TA26 defined the zero point of each channel by the signal on the dark side of the lunar disk,
modeled the tone response as a single power law, $S \propto F^{\Gamma}$, between the JPEG signal
$S$ and the incident flux $F$, with $\Gamma = 0.700$, 0.725, and 0.825 for the red, green, and
blue channels calibrated with field stars, and assumed that vignetting had been removed by the
processing, as indicated by the image metadata. Each of these steps was reasonable given the available data, but none of them could
be tested independently.

\subsection{This work: RAW data as ground truth} \label{sec:intro_thiswork}

On 2026 October~7, the original camera RAW files of the Artemis~II lunar flyby images were
archived in the NASA Planetary Data System \citep{PDSArtemis2CrewCamera}. For
\texttt{art002e009301}, this provides the 14-bit sensor data of exactly the same exposure that
was analyzed in TA26. Such a pairing of a rendered JPEG image and the corresponding RAW data for
a scientifically analyzed astronomical target offers a rare opportunity to check every assumption
of a JPEG-based analysis quantitatively.

The purpose of this paper is twofold. First, we re-derive the properties of the F-corona from the
RAW data with a full reduction, including an in situ flat-field correction, a characterization of
the zero point, and an absolute calibration. A second frame with a three times longer exposure
(\texttt{art002e009300}), reduced independently, provides a key check that the results do not
depend on a single image. Second, we use the RAW
data to evaluate how each step of the JPEG-based analysis affected the results of TA26, and we
assess which quantities can and cannot be derived reliably from consumer-camera JPEG images.
Throughout this paper, ``ground truth'' refers to the closest available representation of the sensor
measurements: the RAW data are themselves stored with a non-reversible compression
(Section~\ref{sec:data_raw}) and are therefore not strictly identical to the original sensor
output.

Section~\ref{sec:data} describes the data, Section~\ref{sec:reduction} the reduction of the RAW
data, and Section~\ref{sec:fcorona} the properties of the F-corona derived from them.
Section~\ref{sec:groundtruth} compares the RAW and JPEG analyses, Section~\ref{sec:capabilities}
discusses the capabilities and limitations of JPEG-based analysis, and
Section~\ref{sec:summary} summarizes this work.

%% =====================================================================
%% Section 2: Data
%% =====================================================================

\section{Data} \label{sec:data}

\subsection{Observation} \label{sec:data_obs} \label{sec:data_second}

\begin{deluxetable*}{lcc}
\tablecaption{Properties of the Analyzed Frames \label{tab:frames}}
\tablehead{
\colhead{Item} & \colhead{\texttt{art002e009301}} & \colhead{\texttt{art002e009300}}
}
\startdata
Role in this work            & Primary (analyzed in TA26)   & Consistency check \\
Exposure start (UTC)         & 2026-04-07 01:06:19.78       & 2026-04-07 01:04:57.46 \\
Exposure time                & 2\,s                          & 6\,s \\
Camera / lens                & \multicolumn{2}{c}{Nikon Z\,9 / AF Nikkor 35\,mm $f/2$D} \\
Aperture / ISO               & \multicolumn{2}{c}{$f/2$ / 1600} \\
RAW format                   & \multicolumn{2}{c}{NEF, 14 bit, HE$^\ast$ compression} \\
RAW size (pixels)            & \multicolumn{2}{c}{$8280 \times 5520$ (RGGB)} \\
Black / saturation level     & \multicolumn{2}{c}{1008 / 15892\,ADU} \\
JPEG size (pixels)           & $8256 \times 5504$ (8 bit, sRGB) & \nodata \\
JPEG rendering               & Lightroom Classic 15.2.1, lens profile & \nodata \\
Field of view                & \multicolumn{2}{c}{$54\fdg0 \times 37\fdg2$} \\
Plate scale                  & \multicolumn{2}{c}{$25''$ (center) -- $20''$ (corners) pixel$^{-1}$} \\
\enddata
\end{deluxetable*}

\begin{figure*}[t]
\centering
\includegraphics[width=\textwidth]{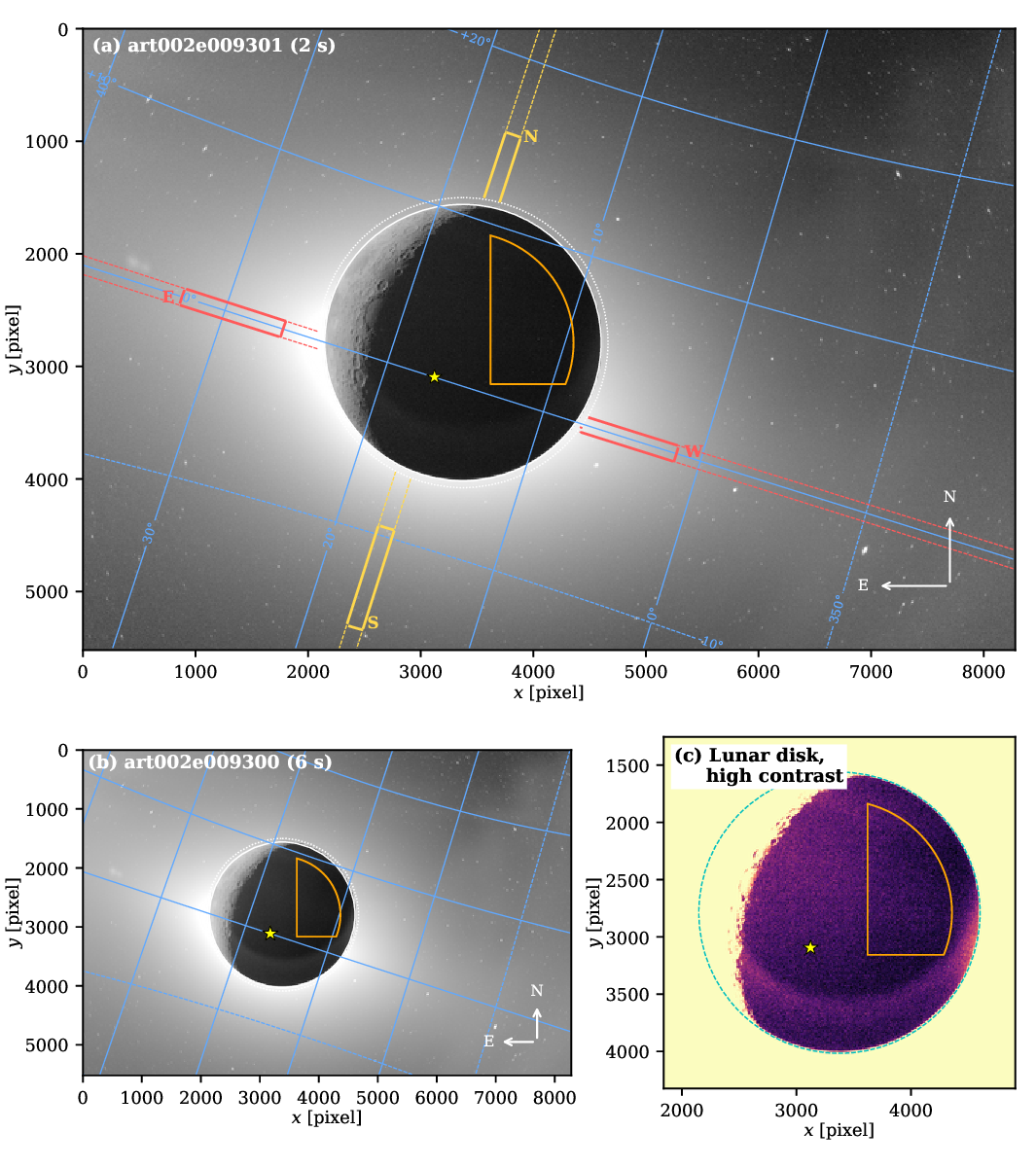}
\caption{(a) RAW image of \texttt{art002e009301} (surface brightness summed over the three channels,
inverse hyperbolic sine stretch) in the pixel coordinates of the RAW sensor array, with north up and
east to the left within $\sim$6$^\circ$. Blue lines show the ecliptic grid at $10^\circ$ intervals, and
the star symbol marks the Sun. White solid and dotted circles indicate the lunar limb and the edge of
the mask, the orange outline the dark-side region used for the zero point, and the red (east and
west) and yellow (north and south) boxes the strips used for the radial profiles (solid: fitted range
of 10$^\circ$--15$^\circ$). (b) Same for \texttt{art002e009300}, without the strips. (c) Lunar disk of
\texttt{art002e009301} with a high contrast, showing the dark-side signal before flat-field correction.
\label{fig:images}}
\end{figure*}

The images were taken by the Artemis~II crew from inside the Orion spacecraft during the solar
eclipse that followed the closest approach to the Moon (Table~\ref{tab:frames};
Figure~\ref{fig:images}). The primary frame, \texttt{art002e009301}, is the image analyzed in TA26.
The second frame, \texttt{art002e009300}, was taken 82\,s earlier with a three times longer
exposure. Between the two exposures, the star field shifted by $\sim$60--80 pixels and rotated by
$\sim$0\fdg5 on the detector, whereas the lunar disk moved by only $\sim$8 pixels, i.e., the camera
was kept pointed at the Moon, which moved relative to the background stars owing to the motion
of Orion. The stellar images of \texttt{art002e009300} are trailed by up to $\sim$10--20 pixels.

\subsection{RAW data} \label{sec:data_raw}

We used the original camera files in Nikon Electronic Format (NEF) from the Artemis~II Crew Camera
Bundle of the NASA Planetary Data System \citep{PDSArtemis2CrewCamera}, rather than the processed
products of the bundle \citep{Artemis2DUG}, so that the entire reduction could be controlled.
The NEF files are recorded with the ``High efficiency$\star$'' (HE$^\ast$) option of the camera, a
non-reversible compression based on the TicoRAW codec \citep{intoPIX2021}, which, according to the
manufacturer, produces pictures that compare favorably in quality to those recorded with lossless
compression \citep{NikonZ9}. We converted the NEF files into the Digital
Negative (DNG) format with Adobe DNG Converter \citep[version 18.7;][]{AdobeDNGConverter}, which decodes the HE$^\ast$ data
without further processing, and read the DNG files with \texttt{rawpy} \citep{rawpy}. The conversion preserves the
undemosaiced sensor data; possible effects of the HE$^\ast$ compression on faint signals could not be
evaluated directly, but the linear response of the stellar signals (Section~\ref{sec:starcal}) and
the agreement of all results between the two frames, whose signal levels differ by a factor of three,
show no evidence of them. We use the
full sensor area (Table~\ref{tab:frames}); the default crop of the DNG files starts at pixel
$(x, y) = (12, 8)$ and matches the size of the released JPEG image.

\subsection{JPEG image} \label{sec:data_jpeg}

For the comparison with TA26, we used the JPEG version of \texttt{art002e009301} released by NASA.
According to its metadata, it was rendered from the RAW file with Adobe Photoshop Lightroom
Classic about four hours after the exposure, using the Adobe lens profile for the lens and camera
with both distortion and vignetting corrections flagged as applied. The coefficients of the
radial distortion model are recorded, and the full lens profile, including a vignetting model, is
also embedded in the DNG files. The parameters of the other rendering steps (tone curve, color
rendering, and any local adjustments) are not recorded. The actual processing is evaluated in
Section~\ref{sec:conversion}.

%% =====================================================================
%% Section 3: RAW Data Reduction
%% Pixel coordinates in this section: zero-based (x, y) of the RAW sensor
%% array, with y increasing in readout order (top-down, "noflip" files).
%% =====================================================================

\section{RAW Data Reduction} \label{sec:reduction}

Pixel coordinates $(x, y)$ refer to the full RAW sensor array, are zero-based, and have $y$
increasing downward; celestial south and west are toward $+y$ and $+x$ within $\sim$6$^\circ$. We
use the normalized radius $\rho = d / d_{\rm max}$, where $d$ is the distance from the image center
$(4139.5, 2759.5)$ and $d_{\rm max} = 4975$ pixels is the half-diagonal of the array.

\subsection{Linear Images} \label{sec:linear}

\begin{figure*}[t]
\centering
\includegraphics[width=0.95\textwidth]{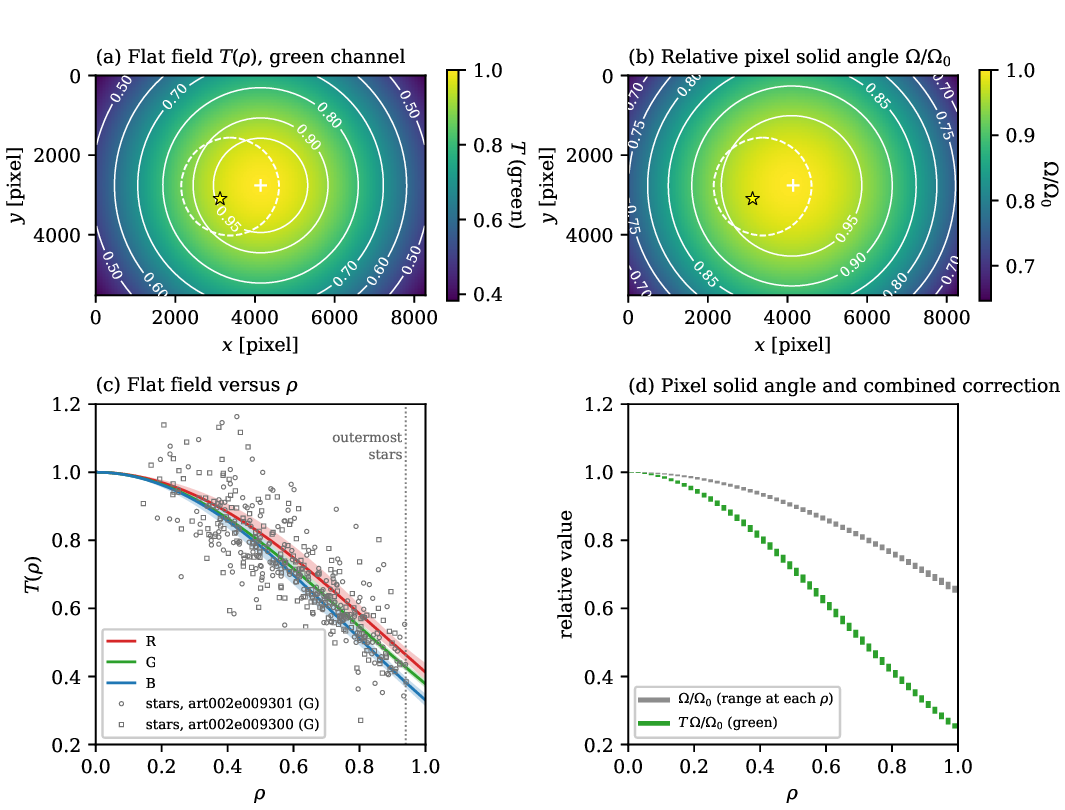}
\caption{Flat field and pixel solid angle of the RAW data.
(a) Flat field $T(\rho)$ of the green channel derived from stellar photometry
(Equations~\ref{eq:vignetting} and \ref{eq:flat}), shown in the pixel coordinates of the RAW sensor array
($y$ increasing downward). The cross marks the image center, the dashed circle the lunar
limb, and the star symbol the position of the Sun in \texttt{art002e009301}.
(b) Relative pixel solid angle $\Omega/\Omega_0$ computed from the astrometric solution.
(c) Flat field versus normalized radius $\rho$ for the three channels ($1\sigma$ bands), with
the green-channel measurements of individual BSC5 stars in \texttt{art002e009301} (circles) and
\texttt{art002e009300} (squares). The dotted line marks the largest $\rho$ of the calibration
stars.
(d) Range of $\Omega/\Omega_0$ at each $\rho$ (gray) and the combined correction for
diffuse emission, $T\,\Omega/\Omega_0$, for the green channel (green; see
Equation~\ref{eq:surface_brightness}).
\label{fig:flat}}
\end{figure*}

After subtraction of the black level, the color filter array data were demosaiced with the
adaptive homogeneity-directed algorithm \citep{Hirakawa2005} of \texttt{LibRaw} \citep{LibRaw} and written as linear images in the
native camera color space, without white balancing, color-space transformation, or tone curve.
\texttt{LibRaw} scales the signal so that the saturation level maps to 65535, i.e., one output unit
corresponds to $1/4.403$\,ADU; signals are quoted in output units below. Saturated pixels were
flagged in the undemosaiced data, and 366 hot pixels were identified as point sources at identical
positions in the two frames despite the shift of the star field between them.

\subsection{Astrometry} \label{sec:astrometry}

An astrometric solution was derived for each frame from the stars of the Yale Bright Star
Catalogue \citep[BSC5;][]{Hoffleit1991}, which is complete to $V \simeq 6.5$. Because of the wide
field and the strong distortion, Simple Imaging Polynomial \citep[SIP;][]{Shupe2005} distortion terms of order 3 or 4
leave residuals of $\sim$45$''$, and we adopted a gnomonic (TAN) projection \citep{Calabretta2002} with
sixth-order SIP terms. The
fits to 239 and 249 stars give rms residuals of $7\farcs2$ and $10\farcs7$ for
\texttt{art002e009301} and \texttt{art002e009300}, respectively, the latter being larger because
of the trailed stellar images. The field of view is $54\fdg0 \times 37\fdg2$, and the plate scale
decreases from $25''$\,pixel$^{-1}$ at the center to $\sim$20$''$\,pixel$^{-1}$ at the corners.

For comparison, the \texttt{Astrometry.net} \citep{Lang2010} solution of the JPEG image used in
TA26, obtained with the default second-order SIP terms, deviates from the BSC5 positions by
59$''$ (rms); because the distortion of the JPEG image has been largely corrected in the rendering
(Section~\ref{sec:conv_geometry}), solving it with SIP orders of 4 or 6 already gives 7--12$''$. This difference is
negligible for the $1^\circ$ bins used in TA26. From the astrometric solution, we computed the
solid angle of each pixel, $\Omega(x, y)$, which is $\Omega_0 = 1.46 \times 10^{-8}$\,sr at the
image center (Figure~\ref{fig:flat}(b)).

\subsection{Position of the Moon} \label{sec:moon}

A circle fitted to the lunar limb of \texttt{art002e009301} has its center at
$(x, y) = (3375.0, 2788.9)$ and a radius of 1227.9 pixels (rms residual of 3.1 pixels),
corresponding to an angular diameter of $16\fdg85$, in agreement with TA26. The fitted radius
gives a distance between Orion and the Moon of 11,860\,km, from which, with the geocentric
ephemerides of the Moon and the Sun, the solar position as seen from Orion is
$(\lambda, \beta) = (16\fdg991, 0\fdg013)$ (J2000 ecliptic coordinates), in agreement with the
position adopted in TA26. The lunar disk was masked with a margin of 60 pixels ($\sim$0\fdg4).

\subsection{Flat Field and Surface Brightness} \label{sec:flat}

Because no flat-field frame is available, the vignetting was determined in situ from aperture
photometry of the BSC5 stars (radius of 32 pixels, local background in an annulus of 40--50
pixels, saturated stars excluded). The difference between the instrumental and catalog magnitudes
was modeled with an even-order polynomial in $\rho$, a form commonly used for the radial light
falloff of camera lenses \citep[e.g.,][]{Goldman2010, AdobeDNG}:
\begin{equation}
  m_{\rm inst} - V = z_k + c\,(B-V) + a_2 \rho^2 + a_4 \rho^4 ,
  \label{eq:vignetting}
\end{equation}
where $z_k$ is a zero point for frame $k$ and $c$ is a color coefficient. Fitting the two frames
jointly (420--430 stars per channel) gives the relative throughput
\begin{equation}
  T(\rho) = 10^{-0.4\,(a_2\rho^2 + a_4\rho^4)}
  \label{eq:flat}
\end{equation}
shown in Figure~\ref{fig:flat}(a) and (c). The light loss in the green channel is $0.25 \pm 0.02$,
$0.57 \pm 0.03$, and $0.92 \pm 0.02$\,mag at $\rho = 0.5$, 0.75, and 0.94, respectively, and is
larger in the blue and smaller in the red channel by $\lesssim$0.12\,mag. Strong vignetting of
this lens at full aperture is also reported in a published lens review, which found a light loss
of 46\% in the image corners on a full-frame sensor \citep{LensTip35f2D}; the exact position of that
measurement is not documented, but the value is comparable to that of our model for a uniformly
illuminated target (green channel, including the pixel solid angle), 48\%--53\% at $\rho = 0.7$--0.75.

The main systematic uncertainty is the dependence on the aperture: the point-spread function
broadens toward the field edges, so that apertures smaller than 32 pixels overestimate the
vignetting, while apertures of 32 and 48 pixels agree within 0.05\,mag at $\rho \leq 0.75$ and
$\sim$0.1\,mag at $\rho = 0.94$. Freeing the position of the optical axis (consistent with the
image center within $\sim$50 pixels), adding a $\rho^6$ term, or fitting the frames separately
changes the flat field by $\lesssim$0.04--0.08\,mag. A fit with the synthetic RGB magnitudes of
\citet{Cardiel2021}, the catalog used in TA26, agrees within 1--2$\sigma$; it requires no color term
but includes only $\sim$45 stars. The calibration stars extend to $\rho = 0.94$; the model is extrapolated only in the
corners of the field, which affect only the outermost profile bins, not used quantitatively.

For diffuse emission, the signal per pixel is in addition proportional to $\Omega$, and the
surface brightness is
\begin{equation}
  I(x, y) \propto \frac{F(x, y)}{T(\rho)\,\Omega(x, y)/\Omega_0} - I_Z ,
  \label{eq:surface_brightness}
\end{equation}
where $F$ is the signal in the linear image and $I_Z$ is the zero point
(Section~\ref{sec:zeropoint}).

\subsection{Zero Point and Stray Light} \label{sec:zeropoint}

Following TA26, we use the dark side of the lunar disk to determine the zero point. The dark-side
region is the part of the disk within $0.8\,R_{\rm M}$ of its center, more than $0.2\,R_{\rm M}$
from it toward $+x$, and less than $0.3\,R_{\rm M}$ from it toward $+y$ ($R_{\rm M}$ is the lunar
radius), which avoids the limb, the earthshine-illuminated eastern part, and a faint brighter band
near the southern limb (Figure~\ref{fig:images}(c)). In \texttt{art002e009301}, the mean signals in
this region are 19, 46, and 50\,ADU in the red, green, and blue channels.

This signal is not mainly an instrumental offset: from \texttt{art002e009301} to \texttt{art002e009300},
it increases by factors of 3.1--3.3, close to the ratio of the effective exposure times measured
from stars (2.95--3.12). It shows no lunar surface features, which disfavors illumination of the
night side of the Moon, and it is bluer than the F-corona, which disfavors F-corona light scattered
within the camera or by the window. After scaling by the exposure-time ratio, its spatial
distribution agrees between the two frames at the same detector position (rms difference of
2.9\,ADU in the green channel) but not when the frames are aligned on the star field (6.8\,ADU),
which excludes sources fixed on the sky. Because the
Moon was almost fixed on the detector, however, light fixed to the camera and light associated with
the Moon cannot be distinguished in this way.

\begin{figure*}[t]
\centering
\includegraphics[width=\textwidth]{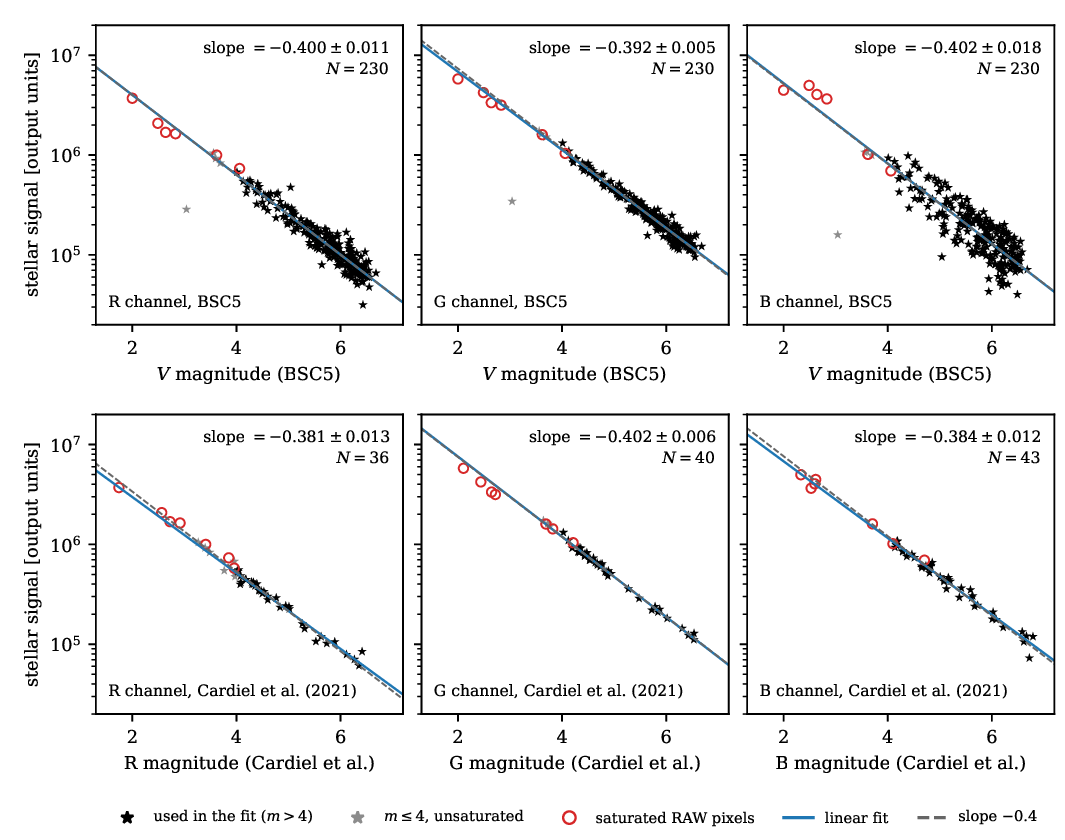}
\caption{Stellar signals in the RAW data of \texttt{art002e009301} as a function of
catalog magnitude for the red, green, and blue channels (left to right). Top: $V$
magnitudes of BSC5. Bottom: synthetic RGB magnitudes of \citet{Cardiel2021}. Black symbols
show the unsaturated stars fainter than 4\,mag used in the fits, gray symbols the
unsaturated brighter stars, and red open circles the stars with saturated RAW pixels. The
signals are measured within a radius of 16 pixels after local background subtraction and
corrected with the flat field. Solid lines show the linear fits, with the slopes and the
numbers of stars given in each panel, and dashed lines show the slope of $-0.4$ expected
for a linear response.
\label{fig:starcal}}
\end{figure*}

A further constraint is provided by the darkest parts of the sky. If the dark-side signal were
uniform on the detector, every sky pixel would exceed it; instead, the mean signals in the
northwestern corner of the field are lower than the dark-side signal by 5--26\,ADU, i.e., by more
than 100 times their uncertainties, so that 3\%--13\% of the sky becomes negative when the dark-side
signal is subtracted uniformly from the image signal. The same comparison in surface brightness,
i.e., after division by $T\,\Omega/\Omega_0$, shows no sky significantly fainter than the dark side
(the faintest sky is 0.97--1.27 times as bright). These observations exclude a detector offset as the main component. Light
confined to the lunar disk, already disfavored above, would also make the near equality of the
faintest sky and the dark side in surface brightness a coincidence. The
observations are naturally explained, although not uniquely, if the dark-side signal is light that enters through the lens and is uniform across the field of view,
such as light scattered in front of the camera (e.g., by the spacecraft window), which is subject to
the same vignetting and pixel solid angle as the sky. We adopt this interpretation and subtract the mean dark-side surface brightness,
\begin{equation}
  I_Z = \left\langle \frac{F}{T\,\Omega/\Omega_0} \right\rangle_{\rm dark} ,
  \label{eq:zeropoint}
\end{equation}
uniformly in surface brightness (Equation~\ref{eq:surface_brightness}). As its systematic
uncertainty, we adopt half the range of $I_Z$ derived from halves of the dark-side region
(2\%--4\%). The two-frame comparison also allows for a constant offset of $-2$ to $-6$\,ADU;
subtracting it separately before the flat-field correction changes the power-law indices of
Section~\ref{sec:profiles} by $\leq 0.07$. For reference, if the dark-side signal were instead uniform on the detector (subtracted before the
flat-field correction), the indices would be steeper by $\leq 0.2$ except toward the south
(0.3--0.8), and subtracting only the black level ($I_Z = 0$) would make them shallower by up to 0.2
along the ecliptic and 0.6 in the north--south direction.

\subsection{Linearity Test with Stars} \label{sec:starcal}

\begin{deluxetable}{lc}
\tablecaption{Error Budget of the Absolute Surface Brightness \label{tab:abserr}}
\tablehead{\colhead{Component} & \colhead{Uncertainty}}
\startdata
Stellar sample (all stars vs.\ $V < 5$) & 5\% \\
Aperture correction & 4\% \\
Zero point and flat field at the profile positions (statistical) & 2\% \\
Flux scale of the $V$ band & 2\% \\
Zero point $I_Z$ (Section~\ref{sec:zeropoint}) at $10^\circ$--$15^\circ$ & $<$0.5\% \\
\hline
Total (quadrature sum) & 7\% \\
\enddata
\tablecomments{For the green channel. The zero point and the flat field are derived from the same
fit to the stars, so that their combination at the positions of the profiles is constrained directly
by the stars there.}
\end{deluxetable}

As a test of the linearity of the reduced data, we compared the stellar signals, measured within
a radius of 16 pixels after local background subtraction and corrected with the flat field, with
catalog magnitudes, as was done for the JPEG image in TA26 (Figure~\ref{fig:starcal}). Because
BSC5 provides no $R$ magnitudes, we used the $V$ magnitudes for all three channels without a color
correction; this hardly affects the slope, the purpose of this test, unless the stellar colors
correlate with magnitude. As a check, we also used the synthetic RGB magnitudes of
\citet{Cardiel2021}, which match the three channels. For
unsaturated stars fainter than 4\,mag, the slopes of $\log_{10} F$ versus magnitude are $-0.392$ to
$-0.402$ with BSC5 and $-0.381$ to $-0.402$ with the synthetic RGB magnitudes of \citet{Cardiel2021},
consistent with the value of $-0.4$ expected for a linear response; stars with saturated RAW pixels
also lie close to the linear relations.

\subsection{Absolute Calibration} \label{sec:abscal}

Because the RAW data are linear, the surface brightness can also be calibrated in absolute units,
which was not possible in TA26. We evaluated the zero point of Equation~(\ref{eq:vignetting}) for the color of the Sun
($B-V = 0.65$), adopted 3636\,Jy for $V = 0$ \citep{Bessell1998}, and applied an aperture
correction of $7\% \pm 4\%$ for the stellar light outside the 32-pixel aperture, derived from the
growth of the signals of bright stars between 16, 32, and 48 pixels. One output unit of the linear
image, corrected with Equation~(\ref{eq:surface_brightness}), then corresponds to
$4.23 \times 10^{-3}$\,MJy\,sr$^{-1}$ in the green channel, which is closest to the $V$ band and to
which we refer below. A calibration based on 48-pixel apertures gives values lower by $\sim$10\%,
because the signals of faint stars in larger apertures are affected by errors in the subtraction of
the bright, structured background; with stars brighter than $V = 5$ only, the two apertures agree
within 5\%, and the values are 4\%--6\% higher than nominal. The resulting error budget is summarized in Table~\ref{tab:abserr}; assuming the components to be
independent, the total uncertainty is $\pm 7\%$. The red and blue channels give
values 4\%--12\% lower, which we regard as a further indication of the systematic uncertainty rather
than a color of the F-corona. The zero point $I_Z$ corresponds to $\sim$0.9\,MJy\,sr$^{-1}$.

%% =====================================================================
%% Section 4: The F-corona from the RAW data
%% =====================================================================

\section{The F-corona from the RAW Data} \label{sec:fcorona}

Below, the surface brightness is that of Equation~(\ref{eq:surface_brightness}) with the zero
point of Equation~(\ref{eq:zeropoint}), calibrated in absolute units as described in
Section~\ref{sec:abscal}. Results refer to \texttt{art002e009301} unless otherwise stated. As in
TA26, east and west refer to increasing and decreasing ecliptic longitude, and north and south to
positive and negative ecliptic latitude.

\subsection{Radial Profiles} \label{sec:profiles}

\begin{figure*}[t]
\centering
\includegraphics[width=\textwidth]{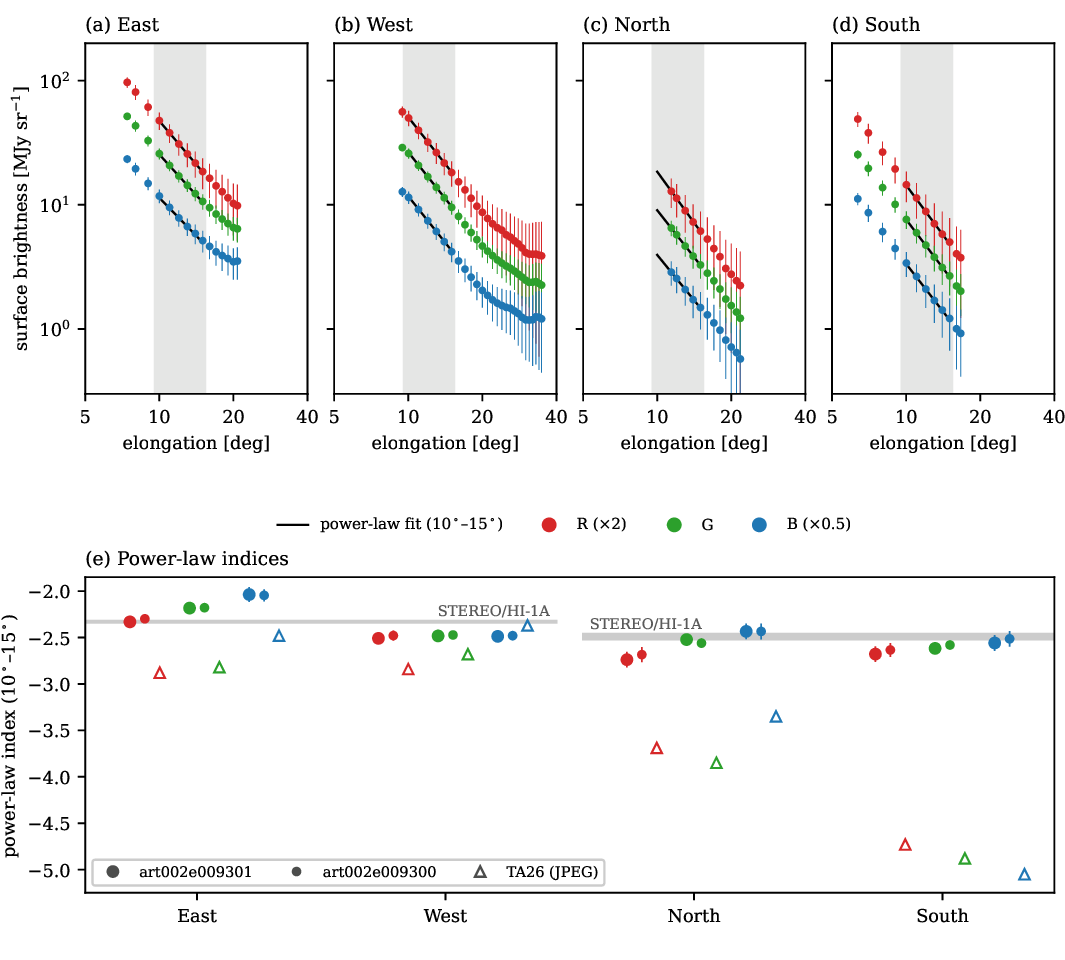}
\caption{(a)--(d) Radial profiles of the surface brightness of the F-corona in
\texttt{art002e009301} toward the east, west, north, and south. The surface brightness is
calibrated in absolute units for each channel as described in Section~\ref{sec:abscal}. Red,
green, and blue symbols show the red, green, and blue channels, multiplied by 2, 1, and 0.5,
respectively, for clarity. Error bars show the standard deviations of the pixel values
in each bin. The shaded areas indicate the elongation range of 10$^\circ$--15$^\circ$
used for the power-law fits, and the solid lines show the fitted power laws. (e) Power-law
indices obtained with the nominal zero point for the four directions and three channels
(Table~\ref{tab:indices}). Large and small circles show the results for
\texttt{art002e009301} and \texttt{art002e009300}, respectively, and open triangles the
JPEG-based results of TA26 (Table~2 therein). The gray bands indicate
the ranges of the indices derived from STEREO/HI-1A along the east--west and north--south
directions \citep{Stenborg2018}.
\label{fig:profiles}}
\end{figure*}

As in TA26, radial profiles were measured along the ecliptic toward the east and west and along the
meridian through the Sun toward the north and south, in strips of $\pm 0\fdg5$ width and $1^\circ$
bins ($3\sigma$-clipped means; Figures~\ref{fig:images}(a) and \ref{fig:profiles}), and fitted with
power laws in the solar elongation $\epsilon$ (the angular distance from the Sun),
$I \propto \epsilon^{p}$, over $10^\circ \leq \epsilon \leq 15^\circ$, where data are
available in all four directions. The uncertainties of $p$ include the fit, the statistical and
systematic uncertainties of the flat field (Section~\ref{sec:flat}), and the uncertainty of the zero
point ($\leq 0.03$).

The indices (Table~\ref{tab:indices}) agree between the two frames within 0.06, although the
exposure times differ by a factor of three. The western profiles have $p \simeq -2.5$ in all
channels, the eastern profiles are shallower ($p = -2.04$ to $-2.33$), and the northern and
southern profiles are similar ($p = -2.43$ to $-2.74$). Beyond $\sim$20$^\circ$, the zero point
amounts to 15\%--50\% of the signal, and the profiles there depend on the assumption that the
foreground light is uniform; we do not use them quantitatively.

\begin{deluxetable*}{llcccc}
\tablecaption{Power-law Indices of the Radial Profiles between Solar Elongations of
$10^\circ$ and $15^\circ$ \label{tab:indices}}
\tablehead{
\colhead{Channel} & \colhead{Frame} & \colhead{East} & \colhead{West} & \colhead{North} & \colhead{South}
}
\startdata
R & \texttt{art002e009301} & $-2.33 \pm 0.04$ & $-2.51 \pm 0.05$ & $-2.74 \pm 0.08$ & $-2.68 \pm 0.08$ \\
  & \texttt{art002e009300} & $-2.30 \pm 0.04$ & $-2.48 \pm 0.05$ & $-2.68 \pm 0.08$ & $-2.63 \pm 0.07$ \\
G & \texttt{art002e009301} & $-2.18 \pm 0.04$ & $-2.48 \pm 0.03$ & $-2.52 \pm 0.05$ & $-2.62 \pm 0.05$ \\
  & \texttt{art002e009300} & $-2.18 \pm 0.04$ & $-2.47 \pm 0.03$ & $-2.56 \pm 0.05$ & $-2.58 \pm 0.05$ \\
B & \texttt{art002e009301} & $-2.04 \pm 0.08$ & $-2.49 \pm 0.05$ & $-2.43 \pm 0.08$ & $-2.56 \pm 0.08$ \\
  & \texttt{art002e009300} & $-2.05 \pm 0.07$ & $-2.48 \pm 0.05$ & $-2.43 \pm 0.08$ & $-2.51 \pm 0.08$ \\
\enddata
\tablecomments{Indices obtained with the nominal zero point (Equation~\ref{eq:zeropoint}).
The uncertainties include the statistical uncertainty of the fit, the statistical and
systematic uncertainties of the flat field, and the systematic uncertainty of the zero
point.}
\end{deluxetable*}

These indices are close to those of previous space-based observations. The mean of the eastern and
western indices ($-2.27$ to $-2.42$) agrees with $-2.31$ to $-2.35$ from STEREO/HI-1A
\citep{Stenborg2018} and $-2.25$ for the equatorial direction in the model of \citet{Koutchmy1985},
and is slightly shallower than $-2.45 \pm 0.05$ from Clementine \citep{Hahn2002}; the northern and
southern indices are comparable to $-2.45$ to $-2.53$ from STEREO/HI-1A and $-2.47$ for the polar
direction in \citet{Koutchmy1985}. An east--west asymmetry in the same sense, although much smaller,
was found with STEREO/HI-1A \citep{Stenborg2018} and Clementine \citep{Hahn2002}. Ours is reproduced
in the two frames, but because the eastern profiles are the most affected by the flat-field
correction (Section~\ref{sec:decomposition}), a residual systematic error of the flat field cannot
be excluded. In contrast, the indices are considerably shallower than those of TA26
(Figure~\ref{fig:profiles}(e)), in particular toward the north and south. As shown in
Section~\ref{sec:decomposition}, this difference arises mainly from the tone response of the JPEG
image, with contributions from the vignetting and the zero point.

\subsection{Color Dependence} \label{sec:color}

Toward the west and south, the indices of the three channels agree within 0.12. Toward the east,
the profiles become shallower from the red to the blue channel, by $+0.29 \pm 0.09$ and
$+0.25 \pm 0.08$ (blue minus red) in the two frames, and a similar but less significant trend is
seen toward the north. The eastern color difference is reproduced in both frames, is present
without the flat-field correction, and is insensitive to the zero point. Systematic effects fixed to
the detector, such as a non-radial component of the vignetting that differs between the channels,
would however affect both frames in the same way, and we regard this color dependence as tentative.
It cannot be explained by contamination from the K-corona: the K-corona has the color of the solar
spectrum, whereas the zodiacal light is redder than the solar spectrum from optical to near-infrared
wavelengths \citep[e.g.,][]{Leinert1998, Tsumura2010}, and the K-corona decreases more steeply with
elongation than the F-corona \citep{vandeHulst1950, Hayes2001}; its contamination would thus steepen
the profiles in the blue channel rather than flatten them.

\begin{figure*}[t]
\centering
\includegraphics[width=\textwidth]{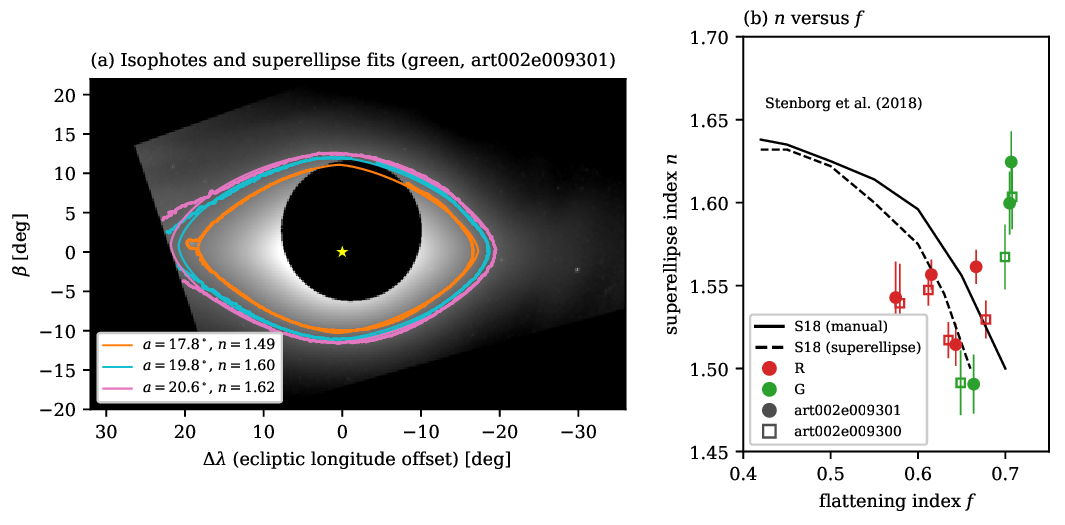}
\caption{(a) Isophotes (dots) of the green channel of \texttt{art002e009301} and the fitted
superellipses (solid curves), shown on a gnomonic grid centered on the Sun (star symbol)
with axes along the ecliptic longitude offset and the ecliptic latitude (east to the left).
The background shows the surface brightness on a logarithmic scale. (b) Superellipse index $n$ versus flattening index $f$ of the fitted
superellipses for the red and green channels of \texttt{art002e009301} (filled circles) and
\texttt{art002e009300} (open squares). The solid and dashed curves show the third-degree
polynomial fits of \citet{Stenborg2018} (their Figure~15) to their measurements obtained
with the manual approach and with superellipse fitting, respectively.
\label{fig:superellipse}}
\end{figure*}

\subsection{Flattening of the Isophotes} \label{sec:flattening}

The flattening of the F-corona was quantified with the flattening index
\citep{Koutchmy1985, Stenborg2018}
\begin{equation}
  f = \frac{R_{\rm eq}}{R_{\rm pol}} - 1 ,
  \label{eq:flattening}
\end{equation}
where $R_{\rm eq}$ and $R_{\rm pol}$ are the equatorial and polar semi-axes of an isophote. From the elongations at which
the four profiles reach a given level, the isophotes with $R_{\rm pol} \simeq 12^\circ$ have
$f_{\rm W} = R_{\rm W}/R_{\rm pol} - 1 = 0.57$--0.60 in all channels and both frames, where the
western semi-axis $R_{\rm W}$ is used because the eastern one reaches the edge of the field.
$f_{\rm W}$ increases outward (0.64--0.77 and 0.78--0.92 for $R_{\rm pol} = 14^\circ$ and $16^\circ$)
and is insensitive to the zero point.

\citet{Stenborg2018} showed that the isophotes of the F-corona between $5^\circ$ and $24^\circ$ are
better described by superellipses than by ellipses. Following their Equation~(4), a superellipse
with center $(x_0, y_0)$, semimajor and semiminor axes $a$ and $b$, and tilt $\psi$ is
\begin{equation}
  \left|\frac{x'}{a}\right|^{n} + \left|\frac{y'}{b}\right|^{n} = 1 ,
  \label{eq:superellipse}
\end{equation}
where $x' = (x - x_0)\cos\psi + (y - y_0)\sin\psi$ and
$y' = -(x - x_0)\sin\psi + (y - y_0)\cos\psi$ are coordinates along the major and minor axes,
and the superellipse index $n$ defines the shape: $n = 2$ gives an ellipse and $n < 2$ a more
diamond-like curve; the flattening index of Equation~(\ref{eq:flattening}) is then $f = a/b - 1$.
\citet{Stenborg2018} found $n$ increasing from $\sim$1.54 to $\sim$1.65 as $f$ decreases from
$\sim$0.66 to $\sim$0.46 toward the Sun. We extracted isophotes from star-removed
surface-brightness images projected onto a Sun-centered gnomonic grid and fitted
Equation~(\ref{eq:superellipse}) with all six parameters free (Figure~\ref{fig:superellipse}). Reliable fits are obtained only for
isophotes with $a \simeq 17^\circ$--$22^\circ$ in the red and green channels, for which we obtain
$n = 1.49$--1.62 and $f = 0.57$--0.71, consistent between the two frames within 0.03. The isophotes
are thus more diamond-like than ellipses ($n = 2$), and the red-channel values lie within
$\sim$0.05 of the relation of \citet{Stenborg2018}. When the centers are fixed at the Sun, the fits are
poorer and $n$ scatters between 1.4 and 2.1 among the isophotes, because the offsets of the isophote
centers (see below) are then absorbed by the shape. We therefore adopt the fits with free centers
and use $n = 1.55 \pm 0.07$, where the uncertainty represents the range among the frames, channels,
and isophotes, as the observed value (Section~\ref{sec:zodi_constant}). The fitted centers are offset
from the Sun by $+0\fdg4$ to $+0\fdg55$ toward the north and by up to $1\fdg2$ toward the east, the
latter reflecting the shallower eastern profiles. The northward offset is consistent with the
viewing geometry discussed by \citet{Stenborg2018}, based on the orientation of the symmetry surface
of the dust cloud \citep{Stenborg2017b}: for an observer between its ascending ($\sim$82$^\circ$) and
descending ($\sim$262$^\circ$) nodes in heliocentric longitude, as in our case ($\sim$197$^\circ$),
the northern part of the cloud is closer and appears brighter.

\subsection{K-corona Contamination} \label{sec:kcorona}

\begin{figure*}[t]
\centering
\includegraphics[width=0.8\textwidth]{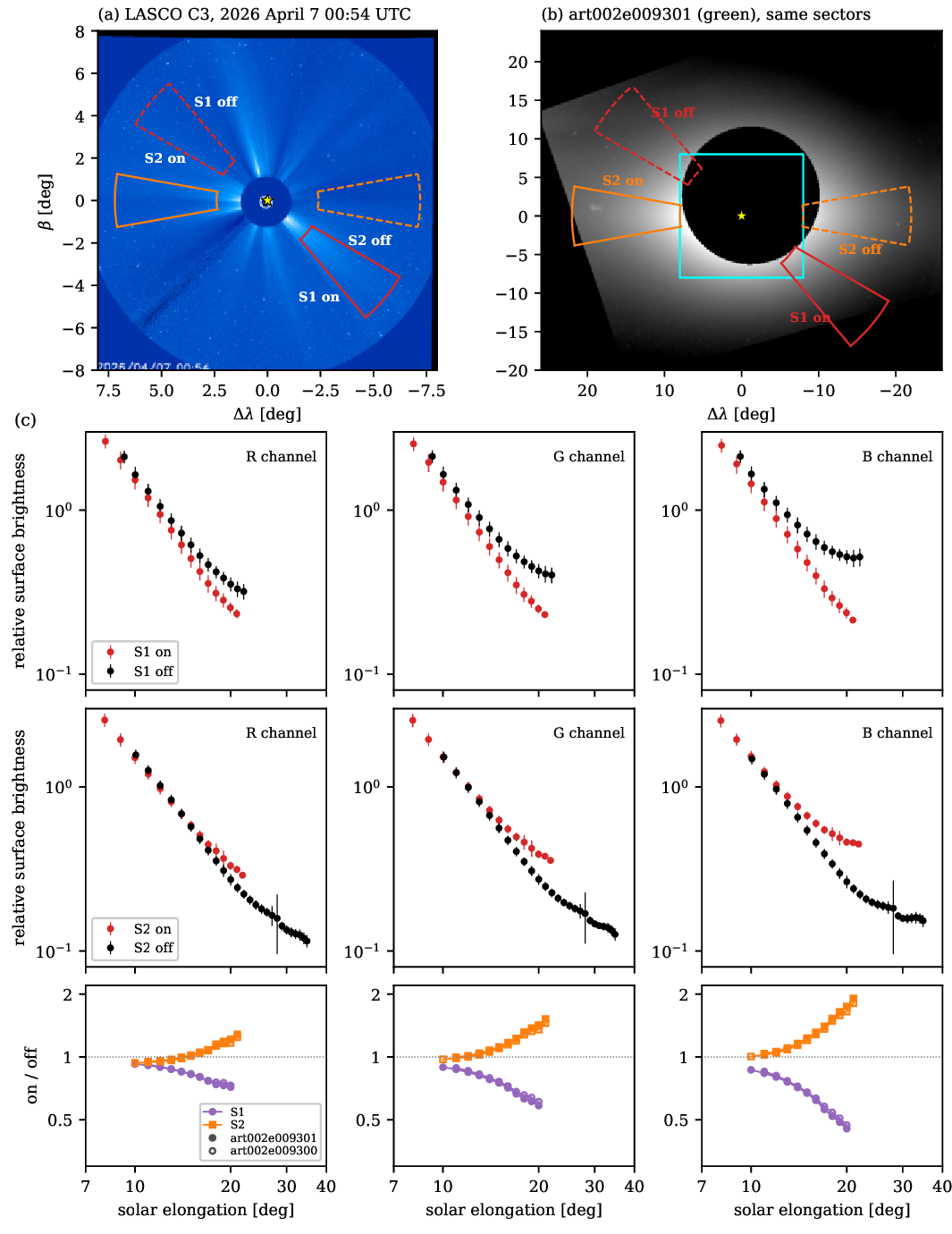}
\caption{Comparison of the brightness along the directions of coronal streamers and along
the opposite directions. (a) LASCO C3 image taken at 00:54 UTC on 2026 April~7, and (b) green
channel of \texttt{art002e009301}, both shown on a gnomonic grid centered on the Sun (star
symbol) with axes along the ecliptic longitude offset and the ecliptic latitude (east to the
left). Solid and dashed outlines show the sectors along the streamers (``on'') and along the
opposite directions (``off''), respectively, for S1 (red) and S2 (orange); the sectors are those of
TA26. The cyan square in panel (b) indicates the field shown in panel (a). (c) Radial profiles for the red, green, and blue channels of
\texttt{art002e009301} (left to right) in the ``on'' (red) and ``off'' (black) sectors of S1
(top row) and S2 (middle row), normalized by the mean of the ``on'' and ``off'' profiles at
12$^\circ$; error bars show the standard deviations within each bin. The ``off'' profile of S1
is shown up to 22$^\circ$, beyond which the sector reaches the corner of the field where the
flat field is extrapolated. The bottom row shows the ratios of the ``on'' to ``off'' profiles
for S1 (circles) and S2 (squares) in \texttt{art002e009301} (filled) and
\texttt{art002e009300} (open).
\label{fig:streamer}}
\end{figure*}

The analyzed elongations ($\epsilon > 7^\circ$, i.e., $\gtrsim$25\,$R_\odot$) are beyond the
distance of $\sim$4--6\,$R_\odot$ at which the K- and F-coronae become comparable
\citep{vandeHulst1950, Hayes2001}, so that the K-corona is expected to be negligible. As a further test, TA26 compared the profiles along two coronal
streamers seen in a publicly available LASCO C3 \citep{Brueckner1995} image taken 12 minutes before
\texttt{art002e009301} with those along the opposite directions. We repeated this comparison with
the same LASCO C3 image and the same sectors as TA26 (Figure~\ref{fig:streamer}). The
ratio of the ``on'' to ``off'' profiles is close to unity at the smallest elongations and departs
from it outward, decreasing to 0.45--0.75 for S1 and increasing to 1.2--1.9 for S2 at
$\sim$20$^\circ$, consistently in the two frames. A K-coronal contribution would instead produce an
excess along the streamers that is largest near the Sun. The departures follow the large-scale
asymmetries found above: the S2 sectors coincide with the eastern and western profiles, and the S1
``off'' sector lies in the northeastern quadrant, toward which the isophotes are displaced. We
therefore find no evidence for a K-coronal contribution, in agreement with TA26, but unlike TA26, we
find a dimming along the streamer only for S1 (see Section~\ref{sec:decomposition}).

\begin{figure*}[t]
\centering
\includegraphics[width=\textwidth]{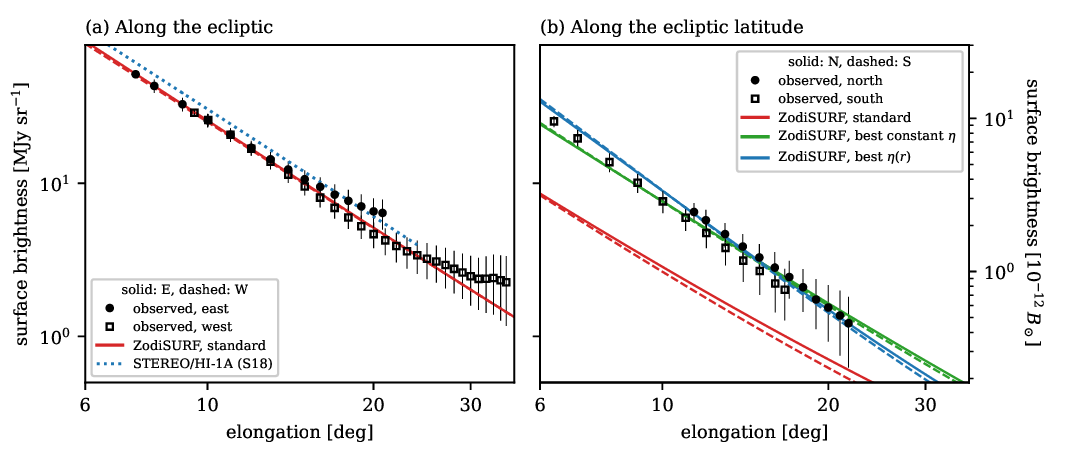}
\caption{Absolute surface brightness of the F-corona in the green channel of
\texttt{art002e009301}, calibrated with field stars (Section~\ref{sec:abscal}). (a) Profiles
along the ecliptic toward the east (filled circles) and west (open squares), compared with the
standard \texttt{ZodiSURF} model (red solid and dashed curves for east and west) and the
STEREO/HI-1A power law of \citet{Stenborg2018} (blue dotted curve). (b) Profiles toward the
north (filled circles) and south (open squares), compared with the standard \texttt{ZodiSURF}
model (red), the best-fitting model with constant vertical-distribution parameters (green), and
the best-fitting model with a distance-dependent vertical distribution (blue; see
Section~\ref{sec:zodisurf}); solid and dashed curves are for the north and south, respectively.
The model curves are for a wavelength of 0.55\,$\mu$m. The right axis gives the surface
brightness in units of $10^{-12}\,B_\odot$. The calibration uncertainty is $\pm 7\%$.
\label{fig:absolute}}
\end{figure*}

\subsection{Absolute Surface Brightness} \label{sec:absolute}

The surface brightness along the ecliptic is $\sim$26\,MJy\,sr$^{-1}$ at $10^\circ$ and
9.5--10.6\,MJy\,sr$^{-1}$ at $15^\circ$, and that toward the north and south is 5.2 and
4.2\,MJy\,sr$^{-1}$ at 12\fdg5 (Figure~\ref{fig:absolute}). For comparison, the power-law fits of
\citet{Stenborg2018} to the eastern photometric axis of the F-corona from STEREO/HI-1A, in units of
the mean solar brightness, give $I \simeq 10^{2.39}\,\epsilon^{-2.33} \times 10^{-11}\,B_\odot$
($\epsilon$ in degrees) for observers between the nodes of the dust cloud, as in our case. With
$B_\odot = 2.66 \times 10^{12}$\,MJy\,sr$^{-1}$ in the $V$ band ($V_\odot = -26.74$), this
corresponds to 30.5 and 11.9\,MJy\,sr$^{-1}$ at $10^\circ$ and $15^\circ$, i.e., 12\%--25\% brighter
than our values, which is consistent given our calibration uncertainty, the different passband of
HI-1, and the different observing geometry.

\subsection{Comparison with a Zodiacal Light Model} \label{sec:zodisurf}

\subsubsection{Model} \label{sec:zodi_model}

Following TA26, we compared the observations with the zodiacal light model \texttt{ZodiSURF}
\citep{OBrien2026}, which combines the dust distribution of \citet{Kelsall1998} with a scattering
phase function and albedo calibrated with HST sky surface-brightness measurements. The density of
the smooth cloud is
\begin{equation}
  n_{\rm d}(r, \zeta) = n_{\rm d,0} \left(\frac{r}{r_0}\right)^{-\alpha} \exp(-\eta\, g^{\gamma}) ,
  \label{eq:kelsall}
\end{equation}
with
\begin{equation}
  g = \begin{cases} \zeta^2 / 2\mu & (\zeta < \mu) \\ \zeta - \mu/2 & (\zeta \geq \mu) \end{cases} ,
  \label{eq:kelsall_g}
\end{equation}
where $r$ is the heliocentric distance, $n_{\rm d,0}$ is the density at $r_0 = 1$\,au, $\zeta$ is the sine of the latitude above the symmetry
plane, and the standard parameters are $\alpha = 1.34$, $\eta = 4.14$, $\gamma = 0.942$, and
$\mu = 0.189$ \citep{Kelsall1998}. We calculated the model surface brightness at
wavelengths of 0.60, 0.53, and 0.47\,$\mu$m for the three channels (0.55\,$\mu$m for the comparison of
absolute surface brightness), for an observer at the Earth at the time of the observation, with the model profiles computed relative to
the solar position used internally by the model. Because the model is calibrated at elongations
$\gtrsim$80$^\circ$, we use it as a phenomenological reference, as in TA26.

\subsubsection{Distribution along the Ecliptic} \label{sec:zodi_ecliptic}

Along the ecliptic, the standard model reproduces both the gradient and the absolute surface
brightness of the F-corona (Figure~\ref{fig:absolute}(a)). The model indices ($-2.31$ and $-2.33$
toward the west and east) agree with the mean of the observed eastern and western indices, and the
predicted surface brightness agrees with the observations within 6\%. The observations therefore support $\alpha \simeq 1.3$ at the heliocentric
distances of $\sim$0.2--0.35\,au that dominate the F-corona at these elongations, as concluded in
TA26. This value agrees with those from zodiacal light observations inside the Earth's orbit,
$\alpha = 1.3 \pm 0.05$ at 0.3--1\,au with Helios \citep{Leinert1981} and $1.30 \pm 0.08$ at
0.76--1.06\,au with Hayabusa2\# \citep{Tsumura2023}, and from STEREO/HI-1A
($1.31$--1.35; \citealt{Stenborg2018}); a single power law with $\alpha \simeq 1.3$ thus describes
the radial distribution of the dust from $\sim$1\,au down to $\sim$0.2\,au. The model is, however,
nearly symmetric between east and west and does not reproduce the shallower eastern profiles.

\subsubsection{Vertical Distribution with Constant Parameters} \label{sec:zodi_constant}

\begin{deluxetable*}{lcccccccc}
\tablecaption{Comparison of \texttt{ZodiSURF} Models with the Observations \label{tab:zodimodels}}
\tablehead{
\colhead{Model} & \colhead{$\eta$ (or $\eta_1$), $\gamma$, $\mu$} & \colhead{$\kappa$} &
\colhead{N--S/E--W} & \colhead{$n$} & \colhead{$f$} & \colhead{$p_{\rm NS}$} & \colhead{$\chi^2$}
}
\startdata
Observed (green channel) & \nodata & \nodata & $0.305 \pm 0.03$ & $1.55 \pm 0.07$ & $0.65 \pm 0.05$ & $-2.60 \pm 0.10$ & \nodata \\
Standard & 4.14, 0.942, 0.189 & 0 & $0.112$ & $1.79$ & $1.84$ & $-2.10$ & $645$ \\
$\eta=2.0$ & 2.0, 0.942, 0.189 & 0 & $0.294$ & $1.71$ & $0.74$ & $-2.25$ & $20.5$ \\
$\gamma=3.0$ & 4.14, 3.0, 0.189 & 0 & $0.331$ & $3.09$ & $0.62$ & $-2.25$ & $499$ \\
$\mu=1.0$ & 4.14, 0.942, 1.0 & 0 & $0.314$ & $2.18$ & $0.68$ & $-2.25$ & $94.3$ \\
Best constant (three observables) & 1.75, 0.85, 0.10 & 0 & $0.303$ & $1.57$ & $0.67$ & $-2.26$ & $12.1$ \\
Best constant (four observables) & 2.25, 0.4, 1.0 & 0 & $0.300$ & $1.51$ & $0.65$ & $-2.26$ & $12.0$ \\
Best $\eta(r)$ & 3.0, 0.942, 0.189 & 0.4 & $0.325$ & $1.64$ & $0.64$ & $-2.65$ & $2.3$ \\
Best $\eta(r)$, $\eta_1$ fixed & 4.14, 0.942, 0.189 & 0.6 & $0.301$ & $1.59$ & $0.69$ & $-2.87$ & $8.2$ \\
\enddata
\tablecomments{N--S/E--W is the ratio of the north--south to the east--west surface brightness at
12\fdg5, $n$ and $f$ are the superellipse and flattening indices of the isophotes with polar radii
of $11^\circ$--$13^\circ$, and $p_{\rm NS}$ is the mean of the northern and southern power-law
indices. $\chi^2$ is computed for all four observables. The best constant models were selected with
the first three observables and with all four observables, respectively.}
\end{deluxetable*}

\begin{figure}[t]
\centering
\includegraphics[width=\columnwidth]{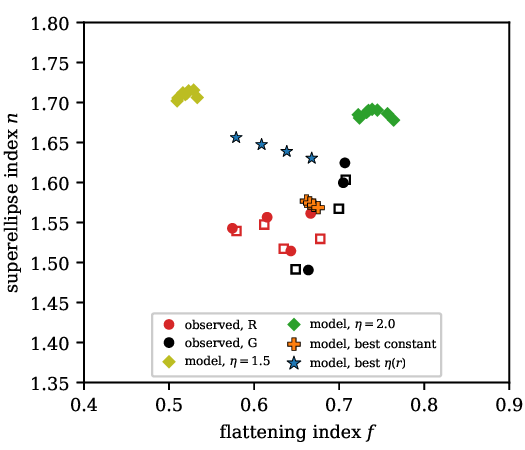}
\caption{Superellipse index $n$ versus flattening index $f$ of the isophotes with semimajor axes
of $\sim$17$^\circ$--22$^\circ$ for the observations (filled circles: \texttt{art002e009301}; open
squares: \texttt{art002e009300}) and for \texttt{ZodiSURF} models, derived with the same method
and field of view. Diamonds show models in which $\eta$ is changed to 1.5 and 2.0; the standard model
and the models with $\eta = 2.5$, $\gamma = 3.0$, and $\mu = 1.0$ lie outside the plotted range
(see Table~\ref{tab:zodimodels}), and the cross and star symbols show the
best-fitting models with constant parameters (Section~\ref{sec:zodi_constant}) and with a
distance-dependent vertical distribution (Section~\ref{sec:zodi_distance}), respectively.
\label{fig:zodisurf}}
\end{figure}

\begin{figure*}[t]
\centering
\includegraphics[width=\textwidth]{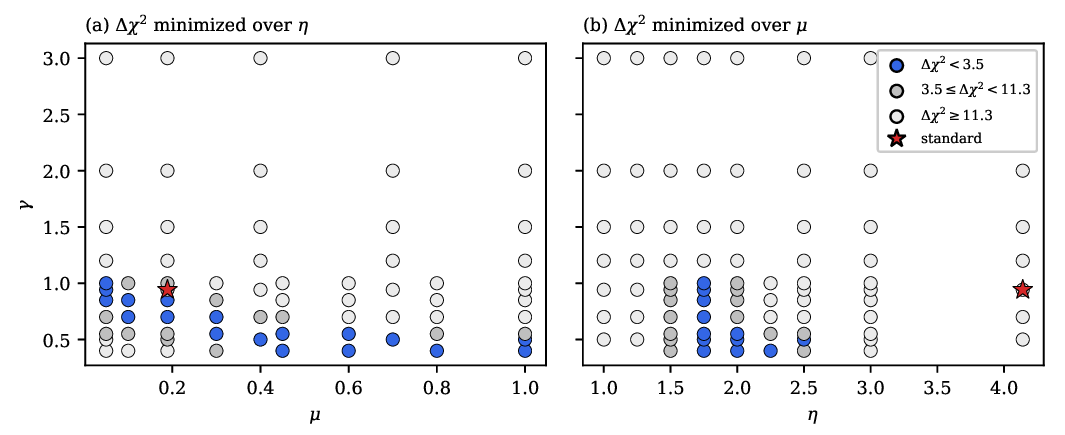}
\caption{Combinations of the vertical-distribution parameters of the \texttt{ZodiSURF} model
(Equations~\ref{eq:kelsall} and \ref{eq:kelsall_g}) compared with the observed north--south to
east--west brightness ratio, superellipse index, and flattening index. Each point represents a
computed combination, shaded according to $\Delta\chi^2$ relative to the best fit after
minimizing over the parameter not shown: (a) $\gamma$ versus $\mu$, minimized over $\eta$;
(b) $\gamma$ versus $\eta$, minimized over $\mu$. Blue points indicate acceptable combinations
($\Delta\chi^2 < 3.5$). The red star marks the standard parameters of \citet{Kelsall1998}.
\label{fig:zodigrid}}
\end{figure*}

Perpendicular to the ecliptic, the standard model underpredicts the surface brightness by a
factor of 2.6--2.9 (Figure~\ref{fig:absolute}(b)), and its flattening ($f \simeq 1.8$) is nearly three
times the observed value. As noted in TA26, the brightness alone does not distinguish between
changes of $\eta$, $\gamma$, and $\mu$, but the shapes of the isophotes respond differently to them
(Figure~\ref{fig:zodisurf}): decreasing $\eta$ mainly reduces the flattening, whereas increasing
$\gamma$ or $\mu$ makes the isophotes much less diamond-like than observed. We therefore computed
the model for a grid of 505 combinations of $\eta$, $\gamma$, and $\mu$ ($\alpha$ fixed) and compared
the ratio of the north--south to the east--west surface brightness at 12\fdg5 and $n$ and $f$ of
the isophotes with polar radii of $11^\circ$--$13^\circ$ with the observed values
(Table~\ref{tab:zodimodels}), using
$\chi^2 = \sum_i [(M_i - O_i)/\sigma_i]^2$, where $O_i$ and $M_i$ are the observed and model values and
$\sigma_i$ the observed uncertainties, which represent the ranges among the frames, channels, and
isophotes and are therefore largely systematic; because the model reproduces the brightness along the
ecliptic, the ratio is equivalent to the absolute brightness perpendicular to it. The acceptable combinations
($\Delta\chi^2 < 3.5$; Figure~\ref{fig:zodigrid}) have $\eta = 1.75$--2.5 and $\gamma \leq 1.0$, with a
trade-off between $\gamma$ and $\mu$ (smaller $\gamma$ requiring larger $\mu$); $\gamma \geq 1.2$ is
excluded. Within this model, the vertical distribution is thus considerably less concentrated toward the
symmetry plane than in the standard model ($\eta \simeq 2$ instead of 4.14). None of these models, however,
reproduces the steeper northern and southern profiles ($p_{\rm NS} \simeq -2.25$, compared with
$-2.43$ to $-2.74$ observed).

\subsubsection{Distance-dependent Vertical Distribution} \label{sec:zodi_distance}

\begin{figure*}[t]
\centering
\includegraphics[width=\textwidth]{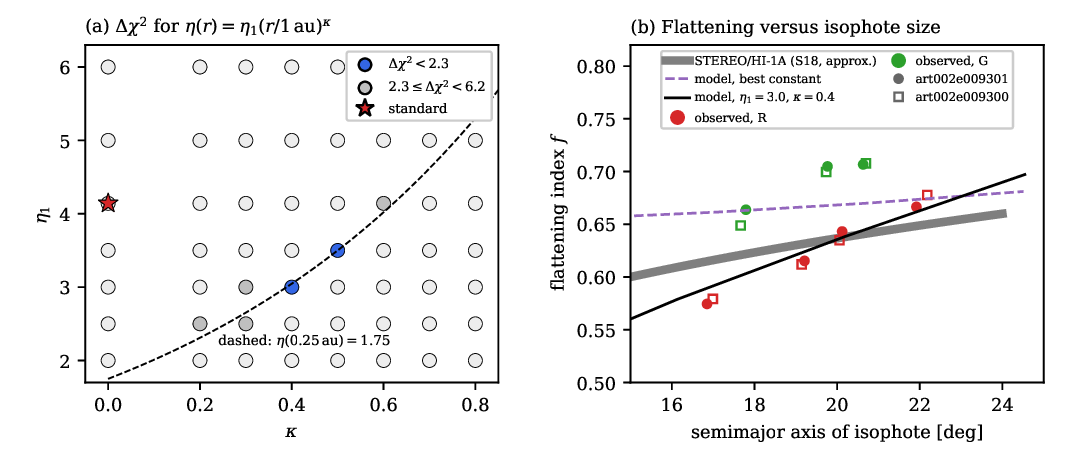}
\caption{Test of a distance-dependent vertical distribution of the dust,
$\eta(r) = \eta_1 (r/1\,{\rm au})^\kappa$ (Equation~\ref{eq:etar}). (a) Combinations of $\eta_1$ and
$\kappa$, with $\gamma$ and $\mu$ fixed at their standard values, shaded according to $\Delta\chi^2$
relative to the best fit for the four observables (north--south to east--west brightness ratio,
superellipse index, flattening index, and northern and southern gradient). The dashed curve
indicates the combinations with $\eta = 1.75$ at $r = 0.25$\,au, and the red star marks the
standard model. (b) Flattening index versus semimajor axis of the isophotes for the
observations (red and green symbols for the red and green channels; filled circles:
\texttt{art002e009301}; open squares: \texttt{art002e009300}), for the best model with constant
parameters (dashed), and for the best model with $\kappa = 0.4$ (solid). The gray band indicates
the approximate trend found by \citet{Stenborg2018} with STEREO/HI-1A; note that their values were
measured with a different method, which gives values larger by $\sim$0.06 than superellipse fits
(their Figure~14).
\label{fig:kappa}}
\end{figure*}

Because the lines of sight at $10^\circ$--$20^\circ$ are dominated by dust at
$r \simeq \sin\epsilon \simeq 0.17$--0.34\,au, whereas the parameters of \citet{Kelsall1998} were
derived from observations at and beyond 1\,au, we tested a vertical distribution that depends on
the distance from the center of the cloud by replacing $\eta$ in Equation~(\ref{eq:kelsall}) with
\begin{equation}
  \eta(r) = \eta_1 \left(\frac{r}{1\,{\rm au}}\right)^{\kappa} ,
  \label{eq:etar}
\end{equation}
so that $\kappa > 0$ makes the cloud less concentrated toward the symmetry plane closer to the Sun.

For a grid of $\eta_1$ and $\kappa$ with standard $\gamma$ and $\mu$, we compared the models with
the three observables above and the mean of the northern and southern indices
($-2.60 \pm 0.10$). The best-fitting model, $(\eta_1, \kappa) = (3.0, 0.4)$, reproduces all four
observables ($\chi^2 = 2.3$), including $p_{\rm NS}$ and the absolute brightness perpendicular to the
ecliptic, while leaving the profiles along the ecliptic unchanged; the best models with constant
parameters give $\chi^2 = 12$ for the same four observables with three free parameters, compared with
two for the distance-dependent model (Table~\ref{tab:zodimodels}; Figure~\ref{fig:kappa}(a)). Because
the uncertainties are largely systematic, these $\chi^2$ values indicate the relative quality of the
fits rather than formal probabilities. With $\eta_1$ fixed at 4.14, the best fit is $\kappa = 0.6$ ($\chi^2 = 8.2$),
with a somewhat too steep $p_{\rm NS}$. The acceptable models have $\eta(0.25\,{\rm au}) \simeq 1.75$,
i.e., the value obtained with constant parameters, while $\kappa \simeq 0.4$--0.6 is constrained by
the northern and southern gradients.

Because these gradients depend on the zero point (Section~\ref{sec:zeropoint}), so does $\kappa$: halving $I_Z$ gives $\kappa \simeq 0.2$--0.3, and subtracting only the black level gives $\kappa \simeq 0$; in contrast, the shapes of the isophotes are unaffected by a uniform zero-point offset, and $\eta \simeq 1.75$--2 with constant parameters is obtained in all these cases.

Moreover, whereas models with constant parameters produce
isophotes with the same flattening at all sizes, the model with $\kappa = 0.4$ predicts a flattening
that increases from $\sim$0.37 at a semimajor axis of $5^\circ$ to $\sim$0.70 at $24^\circ$ and a
superellipse index that decreases from $\sim$1.73 to $\sim$1.62, consistent with the trends found by
\citet{Stenborg2018} ($f \simeq 0.46$ to 0.66 and $n \simeq 1.65$ to 1.54; Figure~\ref{fig:kappa}(b)). Within the framework of
\texttt{ZodiSURF}, the F-corona thus favors a vertical distribution of the dust that broadens toward
the Sun, with $\eta \propto
r^{0.4\text{--}0.6}$ between $\sim$1 and $\sim$0.2\,au, consistent with the decrease of the
flattening of the F-corona toward the Sun \citep{Koutchmy1985, Stenborg2018} and with the efficient
latitudinal scattering of small grains in the inner heliosphere \citep{Ragot2003}.

This does not establish a unique solution:
Equation~(\ref{eq:etar}) is only one possible form, and the vertical distribution cannot be fully
separated from the scattering phase function with a single viewing geometry. Nevertheless, the
F-corona constrains the vertical structure of the dust cloud at $\sim$0.2--0.35\,au, a region not probed by
the infrared observations on which the standard model is based.

\subsection{Limitations} \label{sec:limitations}

The results of this section are subject to the following limitations. First, the zero point relies on
the interpretation of the dark-side signal as foreground light uniform across the field of view
(Section~\ref{sec:zeropoint}); an alternative interpretation would change the indices within $15^\circ$
by $\lesssim$0.2 except toward the south, whereas the profiles beyond $\sim$20$^\circ$ depend on it
strongly. Second, the flat field is derived in situ from stars and is uncertain by up
to $\sim$0.1\,mag at the edge of the field, which may contribute to the east--west asymmetry and the
eastern color dependence. Third, the observation provides a single epoch and viewing geometry, so
that asymmetries due to the observer position cannot be separated from intrinsic ones. Finally, the
constraints on the dust distribution are derived within the \texttt{ZodiSURF} framework, with its
scattering phase function and functional forms, and are therefore model dependent.

%% =====================================================================
%% Section 5: Ground truth for the JPEG analysis
%% =====================================================================

\section{Ground Truth for the JPEG-based Analysis} \label{sec:groundtruth}
\subsection{From RAW to JPEG} \label{sec:conversion}

To verify the results obtained in TA26, we compared the JPEG and RAW images pixel by pixel, after
mapping the pixel grids onto each other.

\subsubsection{Geometric Correction} \label{sec:conv_geometry}

The radial distortion model recorded in the XMP metadata
(Section~\ref{sec:data_jpeg}) reproduces, without any free parameter, the mapping between the
JPEG and RAW pixel grids derived from 1710 stars with an rms residual of 0.40 pixels, the same as
that of a free fifth-order polynomial mapping (0.39 pixels; Appendix~\ref{app:mapping}). The
geometric processing of the JPEG image is thus fully documented in its metadata.

\subsubsection{Vignetting Correction} \label{sec:conv_vignetting}

\begin{figure*}[t]
\centering
\includegraphics[width=\textwidth]{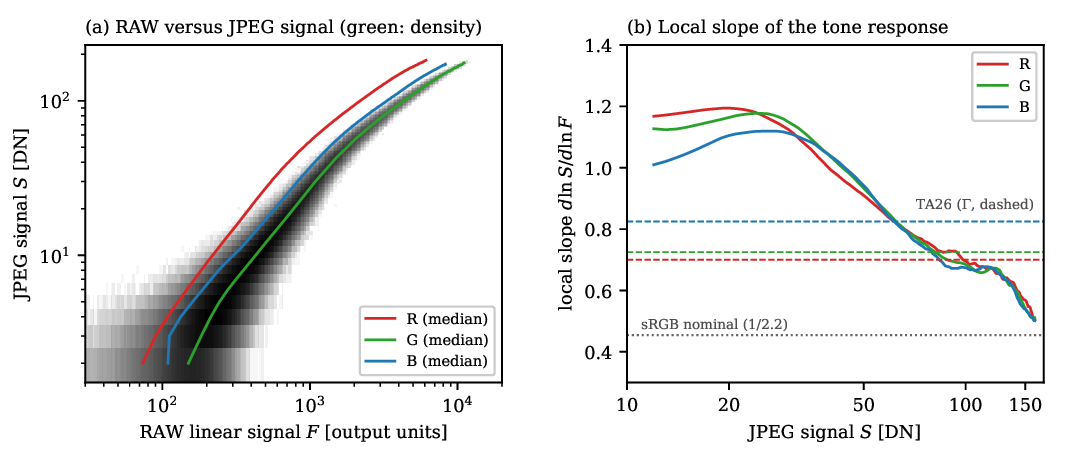}
\caption{Tone response of the JPEG image. (a) JPEG signal $S$ versus linear RAW signal $F$
(before flat-field correction) for matched pixels of \texttt{art002e009301}; the gray scale
shows the density of pixels in the green channel, and the curves show the median RAW signal
at each JPEG level for the three channels. The horizontal offsets between the channels
reflect the white balance applied in the JPEG rendering. (b) Local slope $d\ln S/d\ln F$ of
the median relations. Dashed lines show the values of $\Gamma$ adopted in TA26, and the
dotted line shows the nominal sRGB value of 1/2.2.
\label{fig:tonecurve}}
\end{figure*}

The metadata of the JPEG image indicate that a vignetting correction was applied, and TA26
therefore applied no further correction. However, the vignetting model of the lens profile predicts a light falloff of only 6\% at the image corners, compared with the
actual $\sim$70\% for diffuse light (Section~\ref{sec:flat}). Two tests confirm that the JPEG image
retains essentially the full vignetting. First, if the vignetting had been corrected in the rendering,
the JPEG signal would be a function of $F/T(\rho)$ rather than of the RAW signal $F$, so that pixels
with the same $F$ would have different $S$ depending on $\rho$. Instead, the smoothed JPEG signal is a
tight function of $F$ without any radial correction (rms scatter of 1.5\,DN in the green channel),
whereas assuming a full correction increases the scatter to 5.0\,DN. Second, photometry of the BSC5
stars in the JPEG image, linearized with the tone response derived below, shows a light loss of
0.5--0.7\,mag at $\rho = 0.75$, comparable to that in the RAW data. The assumption of TA26 that the
vignetting had been removed by the processing is therefore not valid.

\subsubsection{Tone Response} \label{sec:conv_tone}

The relation between the JPEG signal $S$ and the linear RAW signal $F$ is tight and monotonic but
is not a single power law (Figure~\ref{fig:tonecurve}). Its local slope $d\ln S/d\ln F$, which
corresponds to $\Gamma$ in TA26, decreases from $\sim$1.1--1.2 at $S \simeq 15$--30\,DN to $\sim$0.7
at $S \simeq 100$\,DN and $\sim$0.5 at $S \simeq 160$\,DN, nearly identically in the three channels,
and neither a single power law nor the sRGB transfer function (approximately $S \propto F^{1/2.2}$)
reproduces it. The values of $\Gamma$
derived in TA26 from stars correspond to the local slopes at $S \simeq 60$--100\,DN, typical of the
backgrounds of the bright calibration stars, and overestimate the compression at the fainter levels
of the outer profiles; their differences among the channels are not present in the actual tone
response. In addition, the JPEG signal does not decrease below $\sim$2--4\,DN, and the JPEG channels
are related to the RAW channels through a color-space transformation that mixes them.

\subsection{Effect of Each Step of the JPEG-based Analysis} \label{sec:decomposition}

\begin{deluxetable*}{lcccccccccccc}
\tabletypesize{\scriptsize}
\tablecaption{Power-law Indices at Each Step of the Analysis \label{tab:decomposition}}
\tablehead{
\colhead{} & \multicolumn{4}{c}{Red} & \multicolumn{4}{c}{Green} & \multicolumn{4}{c}{Blue} \\
\colhead{Step} & \colhead{E} & \colhead{W} & \colhead{N} & \colhead{S} & \colhead{E} & \colhead{W} & \colhead{N} & \colhead{S} & \colhead{E} & \colhead{W} & \colhead{N} & \colhead{S}
}
\startdata
TA26 (Table 2 therein) & $-2.88$ & $-2.84$ & $-3.69$ & $-4.73$ & $-2.82$ & $-2.68$ & $-3.85$ & $-4.88$ & $-2.48$ & $-2.37$ & $-3.35$ & $-5.05$ \\
(1) Method of TA26 & $-2.99$ & $-2.72$ & $-3.97$ & $-4.89$ & $-3.00$ & $-2.61$ & $-4.24$ & $-5.06$ & $-2.60$ & $-2.30$ & $-3.67$ & $-5.43$ \\
(2) Actual tone response & $-2.75$ & $-2.69$ & $-2.80$ & $-3.28$ & $-2.84$ & $-2.69$ & $-2.98$ & $-3.54$ & $-2.81$ & $-2.68$ & $-2.91$ & $-4.23$ \\
(3) + Flat field and pixel solid angle & $-2.21$ & $-2.52$ & $-2.56$ & $-2.80$ & $-2.26$ & $-2.51$ & $-2.71$ & $-3.02$ & $-2.19$ & $-2.50$ & $-2.63$ & $-3.67$ \\
(4) + Zero point in surface brightness & $-2.13$ & $-2.51$ & $-2.48$ & $-2.51$ & $-2.15$ & $-2.49$ & $-2.59$ & $-2.58$ & $-2.01$ & $-2.46$ & $-2.43$ & $-2.78$ \\
RAW data (Table~\ref{tab:indices}) & $-2.33$ & $-2.51$ & $-2.74$ & $-2.68$ & $-2.18$ & $-2.48$ & $-2.52$ & $-2.62$ & $-2.04$ & $-2.49$ & $-2.43$ & $-2.56$ \\
\enddata
\tablecomments{Indices between $10^\circ$ and $15^\circ$ elongation. Steps (1)--(4) are applied to
the JPEG image of \texttt{art002e009301}; see the text.}
\end{deluxetable*}

\begin{figure*}[t]
\centering
\includegraphics[width=\textwidth]{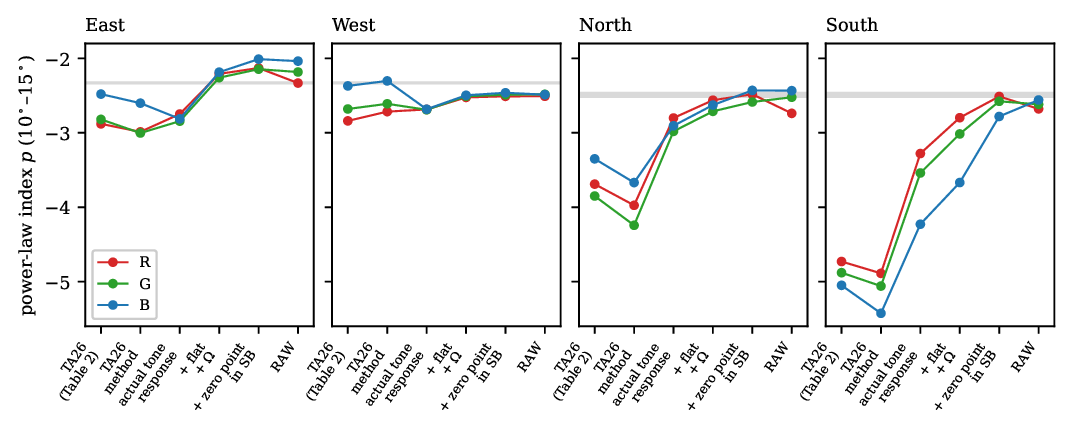}
\caption{Power-law indices of the radial profiles toward the east, west, north, and south at
each step of the analysis (Table~\ref{tab:decomposition}), for the red, green, and blue
channels. The gray bands indicate the ranges of the indices derived from STEREO/HI-1A
\citep{Stenborg2018}.
\label{fig:decomposition}}
\end{figure*}

To isolate the effect of each step, we applied the method of TA26 to the JPEG image and replaced
its steps one by one with those of the RAW analysis (Table~\ref{tab:decomposition};
Figure~\ref{fig:decomposition}): (1) the method of TA26, with their zero point (the sigma-clipped
mean of $S$ in a region on the dark side of the Moon; 2.34, 3.43, and 7.40\,DN) subtracted from $S$, conversion with
$F \propto S^{1/\Gamma}$, and no flat-field correction; (2) conversion with the actual tone response
(Section~\ref{sec:conv_tone}; Figure~\ref{fig:tonecurve}) before the zero-point subtraction; (3) in
addition, correction for the flat field and the pixel solid angle at the corresponding RAW positions
(Section~\ref{sec:flat} and Appendix~\ref{app:mapping}); and (4) in addition, subtraction of the zero
point in surface brightness (Section~\ref{sec:zeropoint}). In all cases, the radial profiles of the
F-corona were measured as in Section~\ref{sec:profiles}. Step (1) differs from TA26 by 0.07--0.39, owing to details of the
implementation such as the lunar mask and the binning.

The largest effect is that of the tone response, which makes the northern and southern profiles
shallower by $\sim$0.8--1.6 and increases the ratio of the north--south to the east--west brightness
at 12\fdg5 from 0.23--0.25 to 0.28--0.30: the northern and southern profiles in the fitted range have
$S \simeq 13$--45\,DN, where the actual local slope (1.0--1.2) is much larger than the adopted
$\Gamma$. The flat-field and pixel-solid-angle correction changes the indices by 0.4--0.6 toward the
east and south and by 0.2--0.3 toward the west and north, reflecting the offset of the Sun from the
optical axis, and the zero point changes the southern indices by 0.3--0.9 and the others by
$\lesssim$0.2. After all corrections, the green-channel indices agree with those from the RAW data
within 0.07; differences of up to 0.26 remain in the red and blue channels, which we attribute to the
mixing of the channels by the color-space transformation.

The same decomposition explains the other differences from TA26. The color dependence of the indices
in TA26 disappears with the actual tone response, i.e., it was caused by the channel-dependent
$\Gamma$. The dimming along both streamers found in TA26 also appears in step (1) (``on'' to ``off''
ratio of 0.74 at 11$^\circ$ for S2 in the green channel) and disappears after the flat-field
correction (1.00), because the eastern ``on'' sector of S2 is vignetted more strongly than the
western ``off'' sector. Similarly, the brightness ratios of the eastern to the western and of the
northern to the southern profiles are 0.65--0.81 and 2.1--3.5 in step (1), compared with 0.95--1.23
and 1.2--1.3 in the RAW data.

\subsection{Assessment of the Conclusions of TA26} \label{sec:ta26check}

\begin{table*}
\centering
\footnotesize
\caption{Assessment of the Conclusions of TA26 with the RAW Data \label{tab:ta26check}}
\begin{tabular}{p{0.42\textwidth}p{0.12\textwidth}p{0.38\textwidth}}
\hline\hline
Conclusion of TA26 & Assessment & RAW result and main cause of the difference \\
\hline
Flattened, nearly elliptical morphology aligned with the ecliptic & Confirmed & Section~\ref{sec:flattening} \\
Flattening index $f = 0.52$--0.56, increasing from red to blue & Revised & $f = 0.57$--0.71 without color dependence; vignetting \\
Northern side brighter than southern side & Confirmed & Weaker (N/S $\simeq 1.2$--1.3); tone response, vignetting \\
Western side brighter than eastern side & Not confirmed & E/W $\simeq 1.0$--1.25; vignetting \\
Power-law profiles steeper than STEREO/HI-1A & Not confirmed & Within $\sim$0.2 of STEREO/HI-1A; tone response, vignetting, zero point \\
Eastern and southern profiles steeper than western and northern & Revised & East shallower than west; north and south similar \\
Significant north--south asymmetry of the indices & Not confirmed & Tone response, zero point \\
Color dependence of the indices (e.g., blue shallowest along the ecliptic) & Not confirmed & Channel-dependent $\Gamma$; a different, tentative eastern color dependence is found \\
K-corona contamination negligible & Confirmed & Streamer dimming partly due to vignetting \\
$\alpha \simeq 1.3$ from comparison with ZodiSURF & Confirmed & Section~\ref{sec:zodisurf} \\
Vertical-distribution parameters degenerate & Partly resolved & $\eta \simeq 1.75$--2.5, $\gamma \lesssim 1.0$ from isophote shapes; $\gamma$--$\mu$ degeneracy remains \\
\hline
\end{tabular}
\end{table*}

\begin{table*}
\centering
\caption{Reliability of Quantities Derived from Consumer-camera JPEG Images \label{tab:reliability}}
\begin{tabular}{p{0.30\textwidth}p{0.14\textwidth}p{0.48\textwidth}}
\hline\hline
Quantity & Reliability & Main limitation and remedy (this work) \\
\hline
Isophote shapes (flattening, superellipse index) & Robust & Invariant under monotonic tone response; verified with RAW data (Table~\ref{tab:decomposition}) \\
Astrometry and geometry & Robust & Requires a high-order distortion model; correction recorded in metadata (Appendix~\ref{app:mapping}) \\
Ordering of signal levels & Robust & e.g., test of the zero-point reference against the darkest sky (Section~\ref{sec:zeropoint}) \\
Radial gradients & Biased & Level-dependent tone response; reconstruct it from stars (Section~\ref{sec:tonerecon}) \\
Comparisons across the field & Biased & Vignetting and pixel solid angle; calibrate with stars and astrometry \\
Faint levels and backgrounds & Biased & Zero point and black level of the rendering; additive constant not recoverable \\
Colors & Biased & Channel-dependent processing and channel mixing \\
Absolute surface brightness & Not available & Requires a linear response and flux calibration \\
\hline
\end{tabular}
\end{table*}

Table~\ref{tab:ta26check} summarizes the main conclusions of TA26 and their assessment. The
flattened morphology is confirmed, and the flattening index from the western semi-axis
($f_{\rm W} = 0.56$--0.59) is the same at all steps of the decomposition, because the shapes of the
isophotes are invariant under a monotonic transformation of the signal. The flattening index using
both semi-axes is, however, reduced by the uncorrected vignetting on the eastern side: with the
normalization and contour levels of TA26, the isophote at a normalized level of 0.1 has $f = 0.49$--0.51
in step (1), comparable to TA26 (0.52--0.56), but $f = 0.62$ in the RAW data (red channel). The steep
radial profiles and the north--south asymmetry of the indices were caused mainly by the tone response,
with contributions from the vignetting and the zero point, and the brighter western side by the
vignetting. The negligible contribution of the K-corona and $\alpha \simeq 1.3$ are confirmed, and
the degeneracy of the vertical-distribution parameters noted in TA26 is partly lifted by the shapes of
the isophotes (Section~\ref{sec:zodisurf}).

%% =====================================================================
%% Section 6: Capabilities and limitations of JPEG-based analysis
%% =====================================================================

\section{Capabilities and Limitations of JPEG-based Analysis} \label{sec:capabilities}

Although our results are specific to one image and one processing chain, the underlying causes are
generic. We summarize them here for analyses of JPEG images from crewed missions, citizen-science
programs, and archives, for which RAW data may not be available.

\begin{figure*}[t]
\centering
\includegraphics[width=\textwidth]{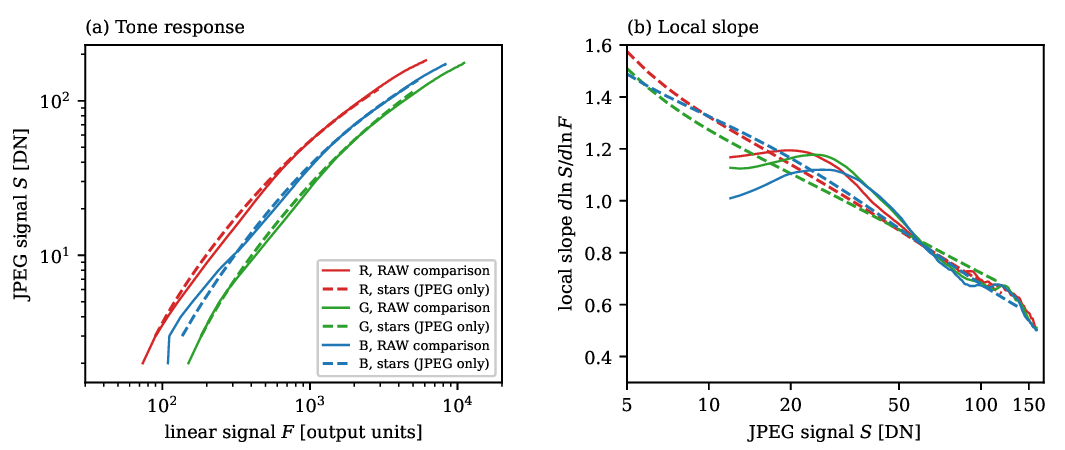}
\caption{Tone response reconstructed from field stars in the JPEG image alone
(Equation~\ref{eq:jpegonly}; dashed curves), compared with that derived from the comparison with
the RAW data (solid curves; Figure~\ref{fig:tonecurve}). (a) JPEG signal versus linear signal.
Because the reconstruction determines the tone response only up to a scale factor and an
additive constant, the reconstructed curves are scaled to agree at $S = 60$\,DN and offset to
agree at $S = 2$\,DN for display. (b) Local slope $d\ln S/d\ln F$. The dashed curves are shown
over the range of background levels covered by the calibration stars.
\label{fig:jpegonly}}
\end{figure*}

\subsection{Robust and Biased Quantities} \label{sec:robust} \label{sec:biased} \label{sec:pitfalls}

Table~\ref{tab:reliability} summarizes which quantities can be derived reliably from consumer-camera
JPEG images, together with their main limitations and the remedies used in this work.
Quantities that depend only on the ordering of the signal are in principle not affected by a
monotonic tone response. The shapes of the isophotes are the prime example, and their robustness is
verified here ($f_{\rm W}$ is the same at all steps of Table~\ref{tab:decomposition}); in general,
however, quantization, compression, channel mixing, and resampling may still affect them, in
particular at faint levels. Other examples are tests that only require knowing which of two signals is brighter, such as the
test of Section~\ref{sec:zeropoint}: that the darkest sky is fainter than the dark side of the Moon,
which shows that the dark-side signal is not uniform on the detector, follows directly from the JPEG
image (1.5--3.0\,DN, compared with 2.4--7.5\,DN) without any knowledge of the tone response.
Geometric quantities are also robust, provided that a distortion model of sufficiently high order is
used, and the geometric correction of the JPEG rendering was fully documented in the metadata.

In contrast, quantities that compare signals at different levels or at different positions on the
detector are biased. In our case, the tone response had the largest
effect, dominating the errors of the northern and southern gradients, followed by the vignetting and
pixel solid angle, which dominated the errors toward the east, and the zero point, which mattered
mainly toward the south (Section~\ref{sec:decomposition}). Field stars, the natural calibrators of a
JPEG image, require care: a star on a bright background measures the local slope of the tone
response at the level of that background, so that a single $\Gamma$ derived from stars is an average
weighted toward the bright backgrounds of the calibration stars; the radial decrease of the
background and the vignetting are partly degenerate when the Sun is near the field center.

\subsection{Reconstruction of the Tone Response from Field Stars} \label{sec:tonerecon}

The tone response can be reconstructed from field stars, because stars on backgrounds of different
brightness probe its local slope at different levels. For a
star of catalog magnitude $m$ and color $B-V$ on a background of JPEG signal $S_B$, the
background-subtracted JPEG signal $\Delta S$ is
\begin{eqnarray}
  \log_{10} \Delta S + 0.4\,m &=& z_{\rm J} + c_{\rm J}\,(B-V) + \log_{10} \frac{dS}{dF}(S_B) \nonumber \\
                              & & - 0.4\,(v_2 \rho^2 + v_4 \rho^4) ,
  \label{eq:jpegonly}
\end{eqnarray}
where $dS/dF$ is the derivative of the tone response, $z_{\rm J}$ is a constant that absorbs the photometric
zero point and the arbitrary normalization of the tone response, $c_{\rm J}$ is a color coefficient, and the
last term is the vignetting with coefficients $v_2$ and $v_4$ (cf.\ Equation~\ref{eq:vignetting}). We fitted this relation to the BSC5 stars in the JPEG image ($\sim$200 per channel, with
$S_B \simeq 2$--130\,DN), representing $\log_{10}(dS/dF)$ as a cubic polynomial in $\log_{10} S_B$, and
integrated it to obtain the tone response up to an additive constant. The reconstructed tone
response agrees with that derived from the RAW data within $\sim$10\% over $S = 15$--150\,DN, and the
derived vignetting agrees within 0.06--0.18\,mag at $\rho = 0.75$ (Figure~\ref{fig:jpegonly}). Using
this calibration with the pixel solid angle and the zero point in surface brightness, and setting the
linear signal to zero at the black level of the JPEG image, reproduces the eastern and western indices
of the RAW data within 0.2 and the north--south to east--west brightness ratio within 0.05, compared
with errors of up to $\sim$3 and 0.07 for the method of TA26. The northern and southern indices still
differ by 0.04--1.0, mainly because the additive constant of the tone response cannot be determined
from the stars: with its true value, the residuals decrease to $\lesssim$0.06 in the red channel and
0.3--0.7 in the green and blue channels, the remainder reflecting the partial degeneracy between the
tone response and the vignetting.

The method is demonstrated here for one image; it requires stars on backgrounds spanning a wide range
of brightness (here 2--130\,DN) and constrains the tone response only over that range; it is applicable to images with
large variations of diffuse brightness, such as the F-corona and the zodiacal light, bright nebulae,
the Milky Way, or skies with strong gradients in twilight or moonlight, but not to images with a
uniformly dark background.

\subsection{Recommendations} \label{sec:recommendations} \label{sec:improvements}

We recommend the following for analyses of consumer-camera JPEG images of diffuse emission when RAW
data are not available.
\begin{enumerate}
\item Inspect the metadata for the rendering software, lens profile, and recorded corrections,
      and do not assume that a flagged correction is complete.
\item Derive the astrometric solution with a distortion model of sufficiently high order
      (e.g., by increasing the SIP order of \texttt{Astrometry.net} from its default value),
      and compute the pixel solid angle from it and the recorded distortion model
      (Appendix~\ref{app:mapping}).
\item Calibrate the tone response as a function of signal level rather than with a single
      $\Gamma$, e.g., with the method of Section~\ref{sec:tonerecon}, and restrict quantitative
      analyses to the range of levels covered by the calibration.
\item Measure the vignetting with field stars, and correct the diffuse emission for both the
      vignetting and the pixel solid angle.
\item Test whether the zero-point reference (e.g., an occulter) contains light that is not
      present over the rest of the field, by comparing it with the darkest parts of the sky.
\item Prefer quantities that are invariant under monotonic transformations, such as isophote
      shapes, for the main conclusions, and treat radial gradients at faint levels, color
      differences, and comparisons across the field with caution.
\end{enumerate}

%% =====================================================================
%% Section 7: Summary
%% =====================================================================

\section{Summary} \label{sec:summary}

We reanalyzed the Artemis~II total solar eclipse images analyzed in TA26, using the RAW sensor
data of the camera. The main results are as follows.

\begin{enumerate}
\item The RAW data were calibrated in situ with field stars. The vignetting of the lens at full
      aperture, combined with the variation of the pixel solid angle, reduces the signal of diffuse
      light by $\sim$70\% at the outer edge of the calibrated field. The signal on the dark side of
      the Moon is light rather than an instrumental offset, and is most naturally explained as
      foreground light that is uniform across the field of view; it was subtracted in surface brightness.
\item Between $10^\circ$ and $15^\circ$ elongation, the radial profiles of the F-corona follow power
      laws with indices of $-2.04$ to $-2.33$ (east), $-2.48$ to $-2.51$ (west), $-2.43$ to $-2.74$
      (north), and $-2.51$ to $-2.68$ (south), consistent with previous space-based observations.
      The isophotes with semimajor axes of $\sim$17$^\circ$--22$^\circ$ have flattening indices of
      $\sim$0.6--0.7 and superellipse indices of $\sim$1.5--1.6. The absolute surface brightness along
      the ecliptic is $\sim$26\,MJy\,sr$^{-1}$ ($\pm 7\%$) at $10^\circ$.
\item Within the framework of the \texttt{ZodiSURF} model, the observations support a radial
      density index of $\alpha \simeq 1.3$ down to heliocentric distances of $\sim$0.2\,au, and
      indicate a vertical distribution of the dust that is much less concentrated toward the symmetry
      plane than in the standard model; a model in which the vertical concentration parameter
      decreases toward the Sun as $\eta \propto r^{0.4\text{--}0.6}$ reproduces all the measured observables, although its gradient depends on the zero point.
\item The JPEG image analyzed in TA26 retains essentially the full vignetting of the lens, and its
      tone response is not a single power law. These two effects, together with the treatment of the
      zero point, account for the steep radial profiles, the north--south asymmetry of the indices,
      and the color dependence reported in TA26.
\item JPEG-based analyses can nevertheless recover robust physical results: the flattened
      morphology, the negligible contribution of the K-corona, and $\alpha \simeq 1.3$ derived in TA26
      survive the RAW analysis, and isophote shapes are robust in JPEG images. Radial gradients, comparisons across the field, and colors, however, require
      calibration of the vignetting and of the tone response (Table~\ref{tab:reliability}); the tone
      response can be reconstructed from field stars when the image contains a wide range of diffuse
      brightness.
\end{enumerate}

These results also bear on future observations of the F-corona from lunar orbit
\citep{Habbal2013, Cooper2023}: foreground light within the instrument field must be characterized,
e.g., by observing with the Moon at different detector positions; a laboratory flat field would
improve on the in situ calibration; and linear data are essential. With RAW data, even a handheld
consumer camera operated by a crew member yields F-corona measurements comparable to those of
dedicated space-based imagers.

%% Please use the acknowledgment and contribution environments. This will 
%% be anonomyized when the "anonymous" style option is used. 
\begin{acknowledgments}
We thank the Artemis~II mission astronauts for capturing this unique image, as well as all members of the Artemis~II mission team who enabled its immediate public release.
The SOHO/LASCO data used here are produced by a consortium of the Naval Research Laboratory (USA), Max-Planck-Institut fuer Aeronomie (Germany), Laboratoire d'Astronomie (France), and the University of Birmingham (UK). SOHO is a project of international cooperation between ESA and NASA.
The authors used Claude (Anthropic), a large language model, to assist with the data analysis and the development of the manuscript,
including the writing and execution of analysis code, the preparation of figures, the drafting and revision of text, improvement of English expression, and refinement of argument structure.
All content generated with AI assistance was critically reviewed, edited, and verified by the authors.
The authors assume full responsibility for the content of the manuscript, including its analyses, interpretations, and conclusions.
\end{acknowledgments}

\begin{contribution}
K.T.\ performed the data analysis and led the writing and submission of the manuscript.
K.A.\ contributed to the data analysis and manuscript preparation.
All authors discussed the results and approved the final manuscript.
\end{contribution}

%\facility{Artemis II}

\software{
\texttt{Adobe DNG Converter} \citep{AdobeDNGConverter},
\texttt{rawpy} \citep{rawpy},
\texttt{LibRaw} \citep{LibRaw},
\texttt{ExifRead} \citep{ExifRead},
\texttt{Astropy} (\citetalias{Astropy2013}, \citeyear{Astropy2018}, \citeyear{Astropy2022}),
\texttt{NumPy} \citep{Harris2020},
\texttt{SciPy} \citep{Virtanen2020},
\texttt{Matplotlib} \citep{Hunter2007},
\texttt{scikit-image} \citep{vanderWalt2014},
\texttt{Astrometry.net} \citep{Lang2010},
\texttt{ZodiSURF} \citep{OBrien2026},
\texttt{IDLAstro} \citep{Landsman1995},
\texttt{JPL Horizons} \citep{Giorgini1996},
\texttt{Claude} \citep{Anthropic2026}
}


\appendix

\section{Geometric Mapping between the JPEG and RAW Images} \label{app:mapping}

The JPEG image was rendered with the radial distortion correction of the Adobe lens profile
recorded in its XMP metadata and in the DNG files (Section~\ref{sec:data_jpeg}). We describe here
the mapping from the pixel coordinates of the JPEG image to those of the RAW sensor array that
reproduces this correction, as used in Section~\ref{sec:decomposition} to evaluate the flat field
and the pixel solid angle at the positions of the JPEG pixels.

We denote by $(x_{\rm J}, y_{\rm J})$ the zero-based pixel coordinates of the JPEG image
($8256 \times 5504$ pixels) and by $(x_{\rm R}, y_{\rm R})$ those of the RAW sensor array
($8280 \times 5520$ pixels; Section~\ref{sec:reduction}), both with $y$ increasing downward. The
centers of the two images are $(c_{x,{\rm J}}, c_{y,{\rm J}}) = (4127.5, 2751.5)$ and
$(c_{x,{\rm R}}, c_{y,{\rm R}}) = (4139.5, 2759.5)$; the offset of $(12, 8)$ pixels corresponds to
the origin of the default crop area of the DNG files. With
$u = x_{\rm J} - c_{x,{\rm J}}$, $v = y_{\rm J} - c_{y,{\rm J}}$, and
$\varrho^2 = (u^2 + v^2)/f_{\rm pix}^2$, the mapping is
\begin{equation}
  \begin{pmatrix} x_{\rm R} \\ y_{\rm R} \end{pmatrix} =
  \begin{pmatrix} c_{x,{\rm R}} \\ c_{y,{\rm R}} \end{pmatrix} +
  s\,\left(1 + k_1 \varrho^2 + k_2 \varrho^4 + k_3 \varrho^6\right)
  \begin{pmatrix} u \\ v \end{pmatrix} ,
  \label{eq:mapping}
\end{equation}
where $k_1 = -0.084258$, $k_2 = 0.086705$, and $k_3 = 0.008748$ are the radial distortion
parameters (\texttt{RadialDistortParam1}--\texttt{3}) and $s = 1.007703$ is the scale factor
(\texttt{ScaleFactor}) of the lens profile, and $f_{\rm pix}$ is the focal length in units of pixels. From the
focal length of the lens (35\,mm) and the focal-plane resolution recorded in the EXIF metadata
(2301.32 pixels\,cm$^{-1}$, i.e., a pixel pitch of 4.345\,$\mu$m), $f_{\rm pix} = 8054.6$ pixels.

We tested Equation~(\ref{eq:mapping}) with 1710 stars detected in both images. Without any free
parameter, it reproduces the measured positions in the RAW image with an rms residual of 0.40
pixels (maximum 1.2 pixels), the same as that of a fifth-order polynomial mapping fitted freely to
the same stars (0.39 pixels). When $f_{\rm pix}$ and $s$ are treated as free parameters, we obtain
$f_{\rm pix} = 8061$ pixels and $s = 1.00769$, in agreement with the recorded values within 0.1\%. Including
the sensor-format factor of the profile (\texttt{SensorFormatFactor} $= 1.003485$) in the
normalization of $\varrho$, or normalizing $\varrho$ after applying $s$, increases the residuals
to 0.41--0.56 pixels.

In Section~\ref{sec:decomposition}, we assumed that the JPEG pixel values are interpolated from the
RAW image at the positions given by Equation~(\ref{eq:mapping}) without rescaling by the change of
pixel area, as in standard image resampling, so that the signal of diffuse emission in a JPEG pixel
is proportional to the solid angle of the RAW pixel at the corresponding position,
$\Omega(x_{\rm R}, y_{\rm R})$, rather than to that of the JPEG pixel. We have not verified this
assumption directly; it is supported by the agreement of the corrected JPEG-based indices with those
from the RAW data within 0.07 in the green channel (Section~\ref{sec:decomposition}), although the
change of pixel area by the distortion correction is only a few percent even at the corners. Under
this assumption, Equation~(\ref{eq:mapping}) and the astrometric solution allow the pixel solid angle
relevant to the JPEG signal to be computed from the metadata alone (Section~\ref{sec:improvements}).

\bibliography{Artemis_ApJ_Tsumura}{}
\bibliographystyle{aasjournalv7}

\end{document}